\documentclass{article}
\usepackage{amssymb,amsmath}
\usepackage{mathrsfs}
\usepackage{amsthm}
\usepackage{hyperref}
\usepackage[dvips]{color}
\usepackage{latexsym} 
\usepackage{graphics}
\usepackage{bm}
\usepackage{color}

\newcommand{\R}{\mathbb{R}} 

\newcommand{\mS}{\mathcal{S}}
\newcommand{\mV}{\mathcal{V}}

\newcommand{\beq}{\begin{equation}}
\newcommand{\eeq}{\end{equation}}

\newcommand{\red}{\textcolor[rgb]{.7,0,0}}
\newcommand{\blue}{\textcolor[rgb]{0,0,.7}}

\newcommand{\diag}{\mathrm{diag}}

\newtheorem{theorem}{Theorem}[section]   
   
\newtheorem{remark}{Remark}[section]

\usepackage{amssymb,amsmath,mathrsfs,mathtools}
\usepackage[dvips]{color}
\usepackage{enumerate}
\usepackage[utf8]{inputenc}
\usepackage[english]{babel}
\usepackage{graphics,graphicx}
\usepackage{latexsym}
\usepackage{times}

\newcommand{\n}{\frac{1}{n}\sum_{i=1}^n}
\newcommand{\argmin}{\mathop{\mathrm{arg\,min}}}
\newcommand{\tr}{\mathrm{tr}}
\newcommand{\tran}{\mathrm{T}}
\newcommand{\Vhh}{\widehat{V}_1(Z)}
\date{}
\begin{document}
\title{Robust and Resistant Regularized Covariance Matrices}
\author{\begin{tabular}{ccccc}  David E. Tyler   & \hspace*{.5cm} & Mengxi Yi& \hspace*{.5cm} & Klaus Nordhausen \\
\normalsize{Department of Statistics} &  & \normalsize{School of Statistics} &  &\normalsize{Dept. of Math.\ and Statistics} \\
\normalsize{Rutgers, The State Univ.\ of NJ} & &  \normalsize{Bejing Normal Univ.\ } & & \normalsize{Univ.\ of Jyv\"{a}skyl\"{a}}\\\normalsize{Piscataway, NJ, U.S.A.}   
& & \normalsize{Beijing, China} & & \normalsize{Jyv\"{a}skyl\"{a}, Finland} \\
\normalsize{dtyler@stat.rutgers.edu}  & & \normalsize{mxyi@bnu.edu.cn} & & \normalsize{klaus.k.nordhausen@jyu.fi} \\
\mbox{  }
\end{tabular}
}

\maketitle

\abstract{\noindent
We introduce a class of regularized M-estimators of multivariate scatter and show, analogous to the popular spatial sign covariance matrix 
(SSCM), that they possess high breakdown points. We also show that the SSCM can be viewed as an extreme member of this class.
Unlike the SSCM, this class of estimators takes into account the shape of the contours of the data  cloud when down-weighing observations. 
We also propose a median based cross validation criterion for selecting the tuning parameter for this class of regularized 
M-estimators. This cross validation criterion helps assure the resulting tuned scatter estimator is a good fit to the data as well as having a
high breakdown point.  A motivation for this new median based criterion is that when it is optimized over all possible scatter parameters, rather than
only over the tuned candidates, it results in a new high breakdown point affine equivariant multivariate scatter statistic.
\\[4pt]
}

\section{Introduction}  
Though being the most efficient at normal distribution, the sample covariance matrix is well known to be non-robust in the sense that it is 
highly influenced  by outliers and has relatively low efficiency at distributions which have longer tails than the multivariate normal distribution. 
A variety of of robust alternatives to the sample covariance matrix have been proposed, with the multivariate M-estimators 
\cite{Maronna:1976, Huber:1977} being one of the earlier ones introduced.  A drawback of the multivariate M-estimators, though, is they are not resistant, 
i.e.\ they have relatively low breakdown points in higher dimensions, specifically no greater than $1/q$, with $q$ being the dimension of the data. 
Consequently, more resistant high-breakdown point estimators of scatter have been developed. These include the minimum volume ellipsoid (MVE) and 
minimum covariance determinant (MCD) estimators \cite{Rousseeuw:1985},  S-estimators \cite{Davies:1987, Lopuhaa:1989}, projection-based scatter estimators
\cite{Donoho-Gasko:1992, Maronna:1992}, and MM-estimators \cite{Tatsuoka-Tyler:2000}. 

All the above estimators are affine equivariant. For a $q$-dimensional sample $X=\{x_1,\ldots, x_n\}$, a multivariate scatter 
statistic, say $V(X)$, is said to be \emph{affine equivariant} if  an affine transformation on the data $x_i \to Ax_i +a$, $i = 1, \ldots, n$, induces 
the  transformation on the scatter statistic  $V(A) \to A V(X) A^{\tran}$, for $a\in\R^q$ and $A\in GL(q)$, the group of 
all $q\times q$ nonsingular matrices. 

As shown in \cite{Davies:1987}, the finite sample contamination breakdown point of any affine equivariant scatter statistics is at most $(n-q+1)/(2n-q+1)$ 
for any data set in general position, which is relatively low when $n$ is not much larger than $q$. Furthermore, unlike the M-estimators, affine equivariant 
high breakdown point estimators tend to be computationally intensive, especially for large $q$. They are typically computed via approximate or 
probabilistic algorithms. To handle these shortcomings, scatter estimators that sacrifice affine equivariance have been proposed, such as the spatial sign 
covariance matrix (SSCM) \cite{Locantore:1999, Visuri:2000}. The SSCM is an appealing candidate because of its simplicity. Not only does it 
have a high breakdown point regardless of $n$, it is also computationally feasible for large $q$.

The SSCM corresponds to a weighted covariance matrix for which the observations are downweighted relative to their Euclidean distance from the center of 
the data, regardless of the shape of the data cloud.  In this paper, we show that a class of regularized M-estimators of scatter, which takes into account the 
shape of the data cloud in downweighing observations, can also have high breakdown points as well as be computationally feasible in high dimensions.  
The SSCM is shown to be an extreme member of this class of regularized scatter estimators, namely as a tuning parameter goes to its upper limit.  The 
interpretation of the SSCM as a regularized M-estimator gives further insight as to why it performs best at spherical distributions, namely since it can be 
viewed as shrinking the scatter statistic towards proportionality to the identity. The broader class of regularized M-estimators of scatter can give improved 
performance when the true scatter matrix deviates from proportionality to the identity. 

The class of regularized M-estimators of scatter only have a high breakdown points for a limited range of the tuning parameters. Also, within this class of 
estimators, there tends to be a trade-off between the breakdown point and how well the estimator can be used to fit the data.  Nevertheless, we propose a 
robust median based cross validation criterion for selecting the tuning parameter which helps assure the resulting tuned scatter estimator is a good fit to the 
data while maintaining a breakdown point close to $50\%$.  A motivation for this new median based criterion is that when it is optimized over all possible
scatter parameters, rather than only over the tuned candidates, it results in a new high breakdown point affine equivariant multivariate scatter statistic, see
Remark \ref{rem:affine}.

The paper is organized as follows. We begin by reviewing the notion of the breakdown point, the definition and properties of elliptical distributions and
 of the spatial sign covariance matrix in Section \ref{Sec:Prelim}.  The definition and properties of penalized M-estimators of multivariate scatter are 
 discussed in Section \ref{Sec:robust}. The results on the breakdown points of these M-estimators for fixed tuning parameters are given in Section \ref{Sec:BP}, 
 wherein a new class of regularized M-estimators of scatter are also introduced. The trade-off between the breakdown point and bias when using fixed 
 turning parameters is discussed in Section \ref{Sec:Bb}. A more detailed looked at tuning is given in Section \ref{Sec:dfM} for the more specific class of 
 distribution-free penalized and regularized M-estimators of scatter. This section also introduces and illustrates the median base cross validation method. 
 Some concluding remarks are made in Section \ref{Sec:conclude}. All formal proofs are given in an appendix.

\section{Preliminary concepts} \label{Sec:Prelim} 
\subsection{Breakdown due to contamination} \label{Sec:bd} 
The breakdown point of an estimator represents a \emph{measure} of its resistance to outliers in the data.
Since its formal introduction in   \cite{Hampel:1971}, a variety of definitions for the breakdown have been proposed. Here we 
use the concept of the finite sample contamination breakdown point, defined in \cite{Donoho:1982,Donoho-Huber:1983}. Roughly translated, 
this corresponds to the proportion of bad data in a sample that can be tolerated by an estimator before making the estimator arbitrarily bad. 
The formal definition is as follows.

Suppose the data set $X=\{x_1,\cdots,  x_n\}$ is contaminated by adding $m$ arbitrary data points $Y=\{y_1,\cdots, y_m\}$. This 
produces an $\epsilon$-contaminated sample $Z = X \cup Y$ consisting of a fraction $\epsilon_m=n/(n+m)$ of arbitrary bad values. Let
$T(\cdot)$ represent a statistic, i.e.\ a function of the data. Next, define an unbounded \emph{bias} function $B(T(X),T(Z))$, which measures
the discrepancy between  $T(X)$ and $T(Z)$ due to the contamination $Y$.  For example, for a location statistic, a natural choice for the 
bias function is $B(t_1,t_2) = \|t_1 - t_2\|$. For $\epsilon=\epsilon_m$, the maximum bias at $X$ caused by $\epsilon$-contamination is 
given by 
\begin{equation} \label{eq:mbias}
b(\epsilon;X)=\begin{cases}
\sup_{Y} B\{T(X),T(Z)\} ,\quad & Z  \in \mathcal{S}_m(X)\\
\infty,\quad & Z \notin \mathcal{S}_m(X)
\end{cases},
\end{equation}
where $\mathcal{S}_m(X)=\{Z \,|\, T(Z) \text{ exists}\}$. Breakdown is said to occur under $\epsilon$-contamination if $b(\epsilon;X)=\infty$.
The finite sample contamination breakdown point of $T(X)$ at $X$ is then defined to be
$\epsilon^*(X;T)=\min_{m \ge 1}\left\{m/(n+m)\,|\,b(\epsilon; X)=\infty\right\}.$

For a $q$-dimensional scatter statistic $V(\cdot)$, i.e. one whose range is the set of positive definite matrices, a natural choice for the bias 
function is the Riemannian metric $B(V_1,V_2) = ||\log(V_1^{-1/2}V_2V_1^{-1/2})||_F$. In this case, breakdown implies either the  
largest eigenvalue of $V(Z)$ can be arbitrarily large, the smallest eigenvalue of $V(Z)$ can be arbitrarily close to zero, or $V(Z)$ does
not exist for some $Y$.

\subsection{Elliptical distributions} \label{Sec:Ellip}
A random vector $x \in \R^q$ is said to have an elliptical distribution if and only if it admits the representation $x = \mu + r \cdot Bu $, with $u$ having a uniform 
distribution on  $\mathcal{S}^{q-1} = \{ u \in \R^q \ | \ u^\tran u = 1 \}$,  $r \ge 0$ being a univariate random variable independent of $u$, $\mu \in \R^q$, 
and $B$ being a fixed $q \times q$ matrix.  The distribution of $x$ depends on $B$ only through $\Sigma_o = BB^\tran$, and so we denote
 $x \sim Elliptical_q(\mu, \Sigma_o; F_r)$ where $r \sim F_r$. Standardizing $z = \Sigma_o^{-1/2} (x -\mu) $ gives a spherical distribution, 
 i.e.\ $=z \sim Qz$ for any orthogonal matrix $Q$, with the distribution of $z$ depending only on the radial distribution $F_r$. This is denoted by 
 $z \sim Spherical_q(F_r)$. 
 
 For $x \sim Elliptical_q(\mu, \Sigma_o; F_r)$, if the first moments of $x$ exists then $\mu$ corresponds to the mean of $X$, and if the second moments
 exist then $\Sigma_o$ is proportional to the covariance matrix of $x$, with the constant of proportionality dependent only on $F_r$. In general, whether
 or not the moments exist, $\mu$ refers to the center of symmetry and $\Sigma_o$ is referred to as the scatter matrix or pseudo-covariance matrix. Note that
 the definition of the scatter matrix is confounded with $F_r$ since $Elliptical_q(\mu, \Sigma_o; F_r) \sim Elliptical_q(\mu, c^2 \Sigma_o; F_{r/c})$.
 However, for an elliptical distribution the \emph{shape} of $\Sigma_o$, defined here as  $\Sigma_{sh,o} \equiv q\Sigma_o/ \tr\{\Sigma_o\}$, is well defined.  
 Furthermore, the shape of $\Sigma_o$ does not depend on $F_r$ and so can be view, together with $\mu$ as well defined parameters within the
 semi-parametric class of elliptical distributions, i.e. without $F_r$ being specified. In general, a shape matrix can be defined as any matrix of the form
$s(\Sigma_o) \Sigma_o$ where $s$ is a positive scalar function such that $s(c^2 \Sigma_o) = s(\Sigma_o)/c^2$. Another common choice is to
take $s(\Sigma_o) = \det(\Sigma_o)^{-1/q}$, see \cite{Davy:2008,Frahm:2009}, in which case the shape matrix then has determinant one rather than trace $q$.

If  $z \sim Spherical_q(F_r)$ is absolutely continuous, its density has spherical contours and is of the form $f_z(z) = g(\|z\|), \ z \in \R^q$, for some 
function $g:\R^+\cup0 \to \R^+\cup0$, where $\|\cdot\|$ refers to the Euclidean norm.  The corresponding density of $x$ has concentric elliptical contours and is given by $f_x(x) = | \det \Sigma_o |^{-1/2} g(\|\Sigma_o^{-1/2} (x -\mu)\|), \ x \in  \R^q$.
For absolutely continuous cases, we denote $z \sim Spherical_q(g)$ and $x \sim Elliptical_q(\mu, \Sigma_o;g)$.

\subsection{Spatial sign covariance matrix} \label{Sec:SSCM}
In this section, we briefly review some properties of spatial sign covariance matrix. For more details, we refer the reader to \cite{Durre:2014},
\cite{Magyar:2014} and \cite{Visuri:2000}.  For $x\in\R^q$, the spatial sign function is defined as $S:\R^q\to\R^q$ with $S(x)=x/\|x\|$
for $x\ne 0$ and $S(0)=0$, where $\|\cdot\|$ denotes the Euclidean norm in $\R^q$. The spatial sign corresponds to the unit vector in the 
direction of $x$. The sample spatial sign covariance matrix is then defined as
\[\mathcal{S}_n(X)=\n S(x_i-\mu_n)S(x_i-\mu_n)^{\tran},\]
where $\mu_n=\argmin_{\mu\in\R^q}\sum_{i=1}^n\|x_i-\mu\|$ is the spatial median. Note that $\n S(x_i-\mu_n)=0$, and so
$\mathcal{S}_n(\cdot)$ corresponds to the sample covariance matrix computed from the spatial signs of the data. The SSCM is scale and location 
invariant and also orthogonally equivariant, i.e. the data transformation $x_i \to c \cdot Q x_i +a$, $i = 1, \ldots, n$  induces  the transformation  
$\mathcal{S}_n(X) \to  Q \mathcal{S}_n(X) Q^{\tran}$, for any $c \in \R$, $a\in\R^q$  and $Q\in O(q)$, the group of orthogonal matrices of order $q$.
The breakdown point of the SSCM is approximately $1/2$. However, if $\mu_n$ is replaced with a fixed location $\mu$, then the breakdown
point becomes one, i.e.\ it cannot break down.

When the data $X$ represents a random sample from $x$ having distribution $F$, then under mild conditions on $F$ the sample SSCM is strongly 
consistent for its functional or population value
$\mathcal{S}(F)=E_F\{S(x-\mu)S(x-\mu)^{\tran}\}$,
where $E_F$ refers to expectation under the distribution $F$,  and $\mu=\mu(F)=\argmin_{\mu\in\R^q}E_F(\|x-\mu\|-\|x\|)$ is the population 
spatial median. When $F$ is an elliptical distribution with finite
second moments, then the SSCM is known to have the same eigenvectors as those of the population covariance matrix, with the corresponding 
eigenvalues having the same order and multiplicities.  However, as shown in \cite{Durre:2016}, the eigenvalues of the population SSCM tend to be
less separated than the eigenvalues of the population covariance matrix. Hence the SSCM can be viewed as shrinking the covariance matrix 
towards proportionality to the identity matrix.

This last property suggests that the SSCM may be related to penalized covariance methods.  In the next section, we review the
penalized M-estimators of multivariate scatter. Their relationship to the SSCM is discussed in Section \ref{Sec:BP}.

\section{Penalized M-estimators of multivariate scatter} \label{Sec:robust}
When one has insufficient data, in that the sample size is small relative to the dimension of the data, regularization or penalization methods are natural to 
consider in order to obtain more stable estimators, see e.g. \cite{Ledoit-Wolf:2004,Yi-Tyler:2021}.
In particular, a ridge approach for regularizing the sample covariance matrices is considered in \cite{Warton:2008}, whereas the existence, 
uniqueness and computation of a regularized Tyler's estimator is studied in \cite{Chen-Wiesel-Hero:2011,Pascal-etal:2014,Sun:2014}. Penalized versions of the 
M-estimators of multivariate scatter were proposed in  \cite{Ollila-Tyler:2014}, with special attention given to penalizing the trace of the precision matrix. For
general penalties, the existence, uniqueness and computation of the penalized M-estimators of scatter are treated in \cite{Duembgen-Tyler:2016}.  

Assume that the q-dimensional data $x_1, \ldots, x_n$ has been robustly centered, e.g.\ by marginal medians or the spatial median, or  
by a known center. Alternatively, the data may be replaced with their pairwise differences. A penalized M-estimator of scatter is then defined as the
matrix $\widehat{\Sigma}$ which minimizes, 
\begin{equation}  \label{eq:M-obj}
L_\rho(\Sigma; \eta) = \frac{1}{n} \sum_{i=1}^n \rho(x_i^{\tran} \Sigma^{-1} x_i) + \log\{ \det(\Sigma)\} + \eta \Pi(\Sigma), \end{equation}
over all $\Sigma > 0$, the set of positive definite symmetric matrices of order $q$. Here $\rho: \R^+ \to \R$, $\Pi(\Sigma)$ denotes a non-negative 
penalty function and $\eta \ge 0$ is the penalty tuning parameter. For $\eta = 0$, \eqref{eq:M-obj} reduces to the objective or loss function associated 
with an M-estimator of scatter based on a given $\rho$-function. Hereafter, we assume $\rho(s)$ is continuously 
differentiable, and let $u(s)=\rho'(s)$, with the ``influence function" $\psi(s)=su(s)$ being non-decreasing.  

We first consider the trace precision matrix penalty $\Pi_{TP}(\Sigma)=\tr(\Sigma^{-1})$, which has the effect of heavily penalizing $\Sigma$ when it is 
nearly singular. Properties of this penalty has been studied in \cite{Ollila-Tyler:2014, Warton:2008}. In particular, it is shown in \cite{Ollila-Tyler:2014}
that, for any $\eta \ge 0$, any critical point $\Sigma>0$ of the penalized loss function \eqref{eq:M-obj} is a solution to the penalized M-estimating 
equation
\begin{equation}\label{eq:OT}
\Sigma=\frac{1}{n} \sum_{i=1}^n u(x_i^{\tran} \Sigma^{-1} x_i)x_ix_i^{\tran}+\eta  I_q.
\end{equation}
Furthermore, if $\rho(s)$ is bounded below, then \eqref{eq:OT} has a unique solution for $\eta > 0$ regardless of the data set. When $\eta = 0$, i.e.\ an
unpenalized M-estimator of scatter, it is known some conditions on the data set are needed to guarantee existence \cite{Kent-Tyler:1991}. 

Similar results hold when using the Kullback-Leibler penalty function $\Pi_{KL}(\Sigma)=\tr(\Sigma^{-1})+\log\det(\Sigma)$. 
Namely, the corresponding M-estimating equation is given by
\begin{equation} \label{eq:KL}
\Sigma=(1-\gamma) \frac{1}{n} \sum_{i=1}^n u(x_i^{\tran} \Sigma^{-1} x_i)x_ix_i^{\tran}+ \gamma  I_q, 
\end{equation}
where $\gamma = \eta/(1+\eta)$. Again,  if $\rho(s)$ is bounded below, then \eqref{eq:KL} has a unique solution for $0 <\gamma \le 1 $ 
regardless of the data set.

If  $u(s) = \kappa/s$ then, as with Tyler's M-estimator \cite{Tyler:1987}, the resulting penalized M-estimator is distribution-free over the class of elliptical 
distributions. This follows since if $x \sim Elliptical_q(0, \Sigma_o; F_r)$ then $x/\|x\| \sim ACG(\Sigma)$, the angular central Gaussian distribution with 
parameter $\Sigma$, see e.g.\ \cite{Tyler:1987b}, which does not depend on $F_r$.  However, the 
corresponding $\rho$-function, given by  $\rho(s) = \kappa \log(s)$, is not bounded below. For this special case, though, it is also shown in 
\cite{Ollila-Tyler:2014} that a unique solution to \eqref{eq:OT} exists regardless of the data whenever $\kappa < 1$. By relating equations \eqref{eq:KL} to 
\eqref{eq:OT}, we see that this statement also applies to \eqref{eq:KL} whenever $(1-\gamma)\kappa < 1$. For other values of $\kappa$, some conditions on 
the sample are needed to ensure existence for these penalized Tyler's M-estimators, see \cite{Ollila-Tyler:2014}.  Since $x/\|x\|$ is not defined for $x = 0$, it is
to be understood that when using the penalized Tyler's M-estimators that the data set $\{x_1, \ldots, x_n\}$ is replaced by the data set consisting only of the 
non-zero data points with corresponding sample size $n_o = \#\{x_i \ne 0; i = 1, \ldots, n\}$.
\begin{remark} \label{Rem:OT-KL}
The class of estimators defined via \eqref{eq:OT} and the class defined via \eqref{eq:KL} are essentially the same class of estimators. Suppose 
$\widehat{\Sigma}_1$ represents the solution to \eqref{eq:OT} with tuning parameter $\eta$ and weight function $u(s) = u_1(s)$ and 
$\widehat{\Sigma}_1$ represents the solution to \eqref{eq:OT} with tuning parameter $\gamma = \eta/(1-\eta)$ and weight function 
$u(s) = u_2(s) = u_1( (1-\gamma)s)$ then we have the relationship $\widehat{\Sigma}_2 =(1-\gamma) \widehat{\Sigma}_1$. 
\end{remark}

\subsection{Scale equivariant versions of the scatter estimates} \label{Sec:scale}
If we wish to pull the scatter estimator towards some $\Gamma > 0$ rather than towards $I_q$, then the penalty term $\Pi(\Sigma)$ can be replaced by 
$\Pi(\Gamma^{-1}\Sigma)$. The resulting estimating equations \eqref{eq:OT} and \eqref{eq:KL} hold provided with $\eta I_q$ and $\gamma I_q$ are replaced 
by $\eta \Gamma$ and $\gamma \Gamma$ respectively.  It is sufficient, though, to only consider the unaltered equations  \eqref{eq:OT} and \eqref{eq:KL}
since the results for general $\Gamma$ can be obtained by applying  \eqref{eq:OT} or \eqref{eq:KL} to the transformed data set 
$y_i = \Gamma^{-1/2} x_i, \ i = 1, \ldots, n$ and noting $\widehat{\Sigma}_X = \Gamma^{1/2} \ \widehat{\Sigma}_Y  \ \Gamma^{1/2}$. 

The solutions to either \eqref{eq:OT} or \eqref{eq:KL} are not in general scale equivariant. That is, if for some $c \in \R$, if we transform the data 
 $x_i \to c \cdot x_i$, $i = 1, \ldots, n$, then the scatter estimator is not simply multiplied by $c^2$.  A scale equivariant scatter estimate can be obtained by 
 choosing $\Gamma = \widehat{\sigma}^2 I_q$, with $\widehat{\sigma}^2$ being an estimate of $\sigma^2$  under the model $\Sigma = \sigma^2 I_q$  
 such that the transformation $x_i \to c \cdot x_i$, $i = 1, \ldots, n$, induces the transformation 
$\widehat{\sigma}^2  \to c^2\widehat{\sigma}^2$. This is similar to the Ledoit-Wolf approach  \cite{Ledoit-Wolf:2004}, which is based on pulling an unrestricted 
estimate of $\Sigma$ towards an estimate based on the model $\Sigma = \sigma^2 I_q$. In their case, they consider only the multivariate normal model 
and so pull the sample covariance matrix $S_n$ towards $s^2 I_q$, with $s^2 = trace(S_n)/q = mean\{(x_i-\bar{x})^\tran(x_i-\bar{x})\}/q$ being
maximum likelihood estimate of $\sigma^2$ under the model $\Sigma = \sigma^2 I_q$.  Rather than use $s^2$ we propose using a robust  
statistic such as $\widehat{\sigma}^2 =  median\{(x_i-\mu_n)^\tran(x_i- \mu_n)\}/q$, with $\mu_n$ being the spatial median. This choice of  
$\widehat{\sigma}^2$ has a breakdown point of $1/2$.

\subsection{Computations}
The penalized M-estimators of scatter  $\widehat{\Sigma}_1$ and $\widehat{\Sigma}_2$ can be computed via simple re-weighing algorithms. Using the 
results for the fixed-point algorithm for finding the solution to \eqref{eq:OT} given in \cite{Ollila-Tyler:2014}, it follows that given any initial starting value 
$\Sigma_0 > 0$, the following algorithms 
 \[ \Sigma_{k+1}=\n u\left(x_i^{\tran} \Sigma_k^{-1} x_i\right)x_ix_i^{\tran}  + \eta I_q \ \mbox{and} \]
 \[ \Sigma_{k+1}=\n (1-\gamma) u\left(x_i^{\tran} \Sigma_k^{-1} x_i\right)x_ix_i^{\tran}  + \gamma I_q \]  
 always converge to  the unique solutions to \eqref{eq:OT} and \eqref{eq:KL} respectively.  For the scale equivariant versions, the terms 
 $\eta  I_q$ and $\gamma I_q$ in the algorithms are replaced by  $\eta  \widehat{\sigma}^2 I_q$ and $\gamma \widehat{\sigma}^2 I_q$ respectively.  
 More generally, if the penalty terms $\Pi_{TP}(\Sigma)$ and $\Pi_{KL}(\Sigma)$ are respectively replaced with $\Pi_{TP}(\Gamma^{-1}\Sigma)$ and 
 $\Pi_{KL}(\Gamma^{-1} \Sigma)$, then  $\eta  I_q$ and $\gamma I_q$ in the algorithms are replaced by  $\eta \Gamma$ and $\gamma \Gamma$ 
 respectively.

\section{Breakdown point results} \label{Sec:BP}
There appears to be little literature regarding the robustness properties of penalized M-estimators of multivariate scatter. In this section, we derive the finite sample contamination breakdown points of the estimator define by \eqref{eq:OT}  and \eqref{eq:KL} above. 

In the following, besides $\rho(s)$ being continuously differentiable, we also presume it is non-decreasing and is either bounded below or 
$\rho(s) = \kappa \log(s)$.  For the latter case, which corresponds to the penalized Tyler's M-estimator of scatter, we further assume $\kappa < 1$ 
when working with \eqref{eq:OT} or $\kappa < 1/(1-\gamma)$ when working with
\eqref{eq:KL}. These conditions assure  \eqref{eq:OT} and \eqref{eq:KL} have unique solutions which we denote by $\widehat{\Sigma}_1$ and
 $\widehat{\Sigma}_2$ respectively. In general, besides $\psi(s) = su(s)$ being non-decreasing, we also presume it is bounded above and 
 define $\kappa = \sup_{s \ge 0} \psi(s) = \psi(\infty)$. 
 We then have the following result, which does not depend on how the data is centered.   The matrix inequalities below refer to the partial ordering of 
 symmetric matrices, i.e.\ $\Sigma \ge \Sigma_o$ if and only if $\Sigma - \Sigma_o \ge 0$. The proof of the following theorem is given in the appendix.  
 
 \begin{theorem} \label{Thrm:bp}
Under the stated assumptions on $u(s)$, regardless of the data  $X = \{x_1,\cdots, x_n\}$,
\[ \eta I_q \le \widehat{\Sigma}_1 \le \frac{\eta}{1-\kappa} I_q, \] 
provided $\kappa < 1$, and 
\[ \gamma I_q \le \widehat{\Sigma}_2 \le \frac{\gamma}{1-(1-\gamma)\kappa} I_q, \]
provided  $\kappa < 1/(1-\gamma)$.
Hence, under these conditions, the estimators $\widehat{\Sigma}_1$ and $\widehat{\Sigma}_2$ cannot break down, i.e.\
 $\epsilon^*(X;\widehat{\Sigma}_1) = \epsilon^*(X;\widehat{\Sigma}_2) = 1$ when $\eta > 0$ and $\max\{0,1-1/\kappa\} < \gamma \le 1$ respectively.
 \end{theorem}
 
 \begin{remark} \label{Rem:scale}
Theorem \ref{Thrm:bp} also holds for the scale equivariant versions, say $\widehat{\Sigma}_{1,\widehat{\sigma}^2}$ and 
$\widehat{\Sigma}_{2,\widehat{\sigma}^2}$, defined in Section \ref{Sec:scale}, provided $I_q$ is replaced by $\widehat{\sigma}^2 I_q$ on both sides of 
the inequalities. Consequently, if $\widehat{\sigma}^2$ has breakdown point $\epsilon^*(X;\widehat{\sigma}^2)$, then it follows that 
$\epsilon^*(X;\widehat{\Sigma}_{1,\widehat{\sigma}^2}) = \epsilon^*(X;\widehat{\Sigma}_{2,\widehat{\sigma}^2}) \ge \epsilon^*(X;\widehat{\sigma}^2)$.
\end{remark}

The low breakdown point of the non-penalized M-estimates of scatter can be partially attributed to the possibility the estimate can be made to be 
arbitrarily close to a singular matrix under certain types of contaminations, see e.g.\ \cite{Duembgen-Tyler:2005}. By contrast, the penalized M-estimators
are obviously bounded away from singularity by construction, resulting in the left hand side of the inequalities in Theorem \ref{Thrm:bp}.
In the following, we show that even if we adjust for this universal bound away from singularity, the resulting regularized estimators still retain high 
breakdown points.  

We first focus on \eqref{eq:KL}.  Define $\widehat{V}_2 = (\widehat{\Sigma}_2 - \gamma I_q)/(1-\gamma)$, where $\widehat{\Sigma}_2$ 
denotes the solution to \eqref{eq:KL}, and hence  $\widehat{V}_2$ is the unique solution to the M-estimating equation
\begin{equation} \label{eq:hyb}
V=\n u\left(x_i^{\tran} \{(1-\gamma) V + \gamma  I_q\}^{-1} x_i\right)x_ix_i^{\tran}
\end{equation}
When $\gamma = 0$, \eqref{eq:hyb} corresponds to the usual M-estimating equations for a scatter matrix, provided a solution exists. On the other hand
when $\gamma = 1$, although $\widehat{\Sigma}_2 = I_q$, we have
\[ \widehat{V}_2 =\n u(x_i^{\tran}x_i)x_ix_i^{\tran}, \]
which corresponds to a weighted covariance matrix with the data points being down-weighted based on their Euclidean distance from the center.
These weighted covariance matrices have recently been studied in \cite{Rousseeuw:2019}, wherein they are referred to as general spatial sign 
covariance matrices.  Note that for $u(s)=1/s$ one obtains the usual SSCM. 
  
Consider now the solution to \eqref{eq:OT},  denoted by $\widehat{\Sigma}_1$, and define 
$\widehat{V}_1 = \widehat{\Sigma}_1 - \eta I_q$, which is the unique solution to the M-estimating equation:
\begin{equation} \label{eq:hyb2}
V=\n u\left(x_i^{\tran} (V + \eta  I_q)^{-1} x_i\right)x_ix_i^{\tran}.
\end{equation}
When $\eta = 0$, this corresponds to  the usual M-estimating equations for a scatter matrix, provided a solution exists. The relationship between 
\eqref{eq:hyb2} and the general SSCM is more involved.  Suppose we consider the family of weight functions $u(s) = w(\eta s)$ for a fixed 
function $w(\cdot)$. Then, as $\eta \to \infty$, we have $\widehat{V}_1 \to \n w(x_i^{\tran}x_i)x_ix_i^{\tran}$.

Rather than choose either extreme as one's estimator,  \eqref{eq:hyb} and \eqref{eq:hyb2} provide classes of scatter 
estimators with tuning parameters $0 \le \gamma \le 1$ and $\eta \ge 0$ respectively. 
Moreover, the following theorem shows these adjusted regularized M-estimators have the same breakdown properties as the SSCM. 
The choice of the tuning parameter is discussed in a later section.

\begin{theorem} \label{Thrm:bp2}
In addition to the stated assumptions on $u(s)$, suppose $u(s) > 0$ is also non-increasing.
Let $\widehat{V}_1 = \widehat{\Sigma}_1 - \eta  I_q$ and $\widehat{V}_2 = (\widehat{\Sigma}_2 - \gamma I_q)/(1-\gamma)$,
with $\widehat{\Sigma}_1$ and  $\widehat{\Sigma}_2$  being defined as in Theorem \ref{Thrm:bp}.  Furthermore,
suppose the data  $X=\{x_1, \cdots, x_n\}$ spans $\R^q$, and is centered by a known or fixed center. Then, 
\begin{itemize}
\item[i)] for  $\eta > 0$, \ $\epsilon^*(X;\widehat{V}_1) = 1$ \ provided \ $\kappa < 1$. 
\item[ii)] for  $\max\{0, (\kappa - 1)/\kappa\} < \gamma \le 1$,  \ $\epsilon^*(X;\widehat{V}_2) = 1$.
\end{itemize}
\end{theorem}

\begin{remark} \label{Rem:center}
If the data is centered by a location statistic $\widehat{\mu}$ having breakdown point $\epsilon^*(X;\widehat{\mu})$, then it can be shown that under the
conditions of Theorem \ref{Thrm:bp2} that
$\epsilon^*(X;\widehat{V}_1) = \epsilon^*(X;\widehat{V}_2) \ge \epsilon^*(X;\widehat{\mu})$.
\end{remark}

\begin{remark} \label{Rem:scale2}
For the scale equivariant versions $\widehat{\Sigma}_{1,\widehat{\sigma}^2}$ and $\widehat{\Sigma}_{2,\widehat{\sigma}^2}$, defined
$\widehat{V}_{1,\widehat{\sigma}^2} = \widehat{\Sigma}_{1,\widehat{\sigma}^2}- \eta \widehat{\sigma}^2 I_q$ and 
$\widehat{V}_{2,\widehat{\sigma}^2} = (\widehat{\Sigma}_{1,\widehat{\sigma}^2} - \gamma \widehat{\sigma}^2 I_q)/(1 - \gamma)$, which are also
scale equivariant. Under the conditions of Theorem \ref{Thrm:bp2}, it can then be shown that 
$\epsilon^*(X;\widehat{V}_{1,\widehat{\sigma}^2}) = \epsilon^*(X;\widehat{V}_{2,\widehat{\sigma}^2}) \ge 
\min\{\epsilon^*(X;\widehat{\mu}), \epsilon^*(X;\widehat{\sigma}^2)\}$. 
\end{remark}

\begin{remark} \label{Rem:bp}
A more detail look at the breakdown point for general $\kappa$ is discussed in Section \ref{Sec:dfM} for the special case $u(s) = \kappa/s$. 
\end{remark}

\section{Bias versus breakdown considerations} \label{Sec:Bb}
Partially due to the simple relationship of $\widehat{V}_2$ to the general SSCM, we hereafter focus on the scatter statistics $\widehat{\Sigma}_2$ and 
$\widehat{V}_2$, Analogous results for $\widehat{\Sigma}_1$ and $\widehat{V}_1$ can be obtained by using their relationship to $\widehat{\Sigma}_2$
and $\widehat{V}_2$; see Remark \ref{Rem:OT-KL}.  Also, we drop the subscript for $\widehat{\Sigma}_2$ and $\widehat{V}_2 $ and simply 
denote them as  $\widehat{\Sigma}$ and $\widehat{V}$. 

When using robust estimators, a natural question to address is what is being estimated, and the general answer is the corresponding functional.
For a given $\gamma$, the functional versions of $\widehat{\Sigma}$ and $\widehat{V}$ can be defined for the former as the solution $\Sigma(F) > 0$ 
to the matrix equation
 \begin{equation} \label{eq:func} \Sigma = (1-\gamma) E_F[u(x^\tran \Sigma^{-1} x) xx^\tran] + \gamma I_q, \end{equation} 
 where $x \in \R^q$ is a random vector with distribution $F$. Again, we assume the function $u(s)$ satisfies the conditions discussed in 
 Section \ref{Sec:robust}, which assures \eqref{eq:func} has a unique solution. The functional version of $\widehat{V}$ is then given by
 $V(F) = (\Sigma(F) - \gamma I_q)/(1-\gamma)$.
 
For a parametric class of distributions, the functionals themselves can be expressed in terms of the parameters. For elliptical distributions,
the following theorem establishes an important property $\Sigma(F)$ and $V(F)$ share with the functional version of the SSCM, namely they have 
have the same eigenvectors as $\Sigma_o$ as well as the same ordering and multiplicities of their eigenvalues.

\begin{theorem}\label{Thrm:eigen}
For $F \sim Elliptical_q(0,\Sigma_o;F_r)$ and with $u(s)$ being non-increasing, the matrices $\Sigma(F)$, $V(F)$ and $\Sigma_o$ can be simultaneously 
diagonalized, i.e.\ for some orthogonal matrix $P$, 
 \[ \Sigma_o=P \Lambda_o P^\tran, \Sigma(F) = P \Lambda_F P^\tran, \text{ and } V(F) = P \Lambda_{v,F} P^\tran, \]
where $\Lambda_o=diag\{\lambda_{o,1},\cdots,\lambda_{o,q}\}$ and $\Lambda=diag(\lambda_1,\cdots,\lambda_q)$, with 
$\lambda_{o,1} \ge \cdots \ge \lambda_{o,q}$ and $\lambda_1 \ge \cdots \ge \lambda_q$. Also,
$\Lambda_{v,F} =  diag(\lambda_{v,1}, \cdots, \lambda_{v,q})$, with $\lambda_{v,j} = (\lambda_j - \gamma)/(1-\gamma), \  j = 1, \ldots, q$.
Furthermore, for $1\le j \le q-1$, $\lambda_{o,j}=\lambda_{o,j+1}\Leftrightarrow\lambda_{j}=\lambda_{j+1}\Leftrightarrow\lambda_{v,j}=\lambda_{v,j+1}$,
 and  if  $\lambda_{o,j}> \lambda_{o,j+1}$ then $\lambda_j > \lambda_{j+1}$ and $\lambda_{v,j}> \lambda_{v,j+1}$. 
\end{theorem}
\noindent
The elements of $\Lambda_F$, and hence of $\Lambda_{v,F}$, are strictly functions of the elements of $\Lambda_o$. In particular, as shown within the proof of
Theorem \ref{Thrm:eigen}, $(\lambda_1, \ldots, \lambda_q)$ is the unique solution to the system of equations
\begin{equation} \label{eq:Lambda}
\lambda_j = (1-\gamma) \lambda_{o,j} E\left[ \ u\left( \ \sum_{k=1}^q  \frac{\lambda_{o,k} \ Z_k^2}{\lambda_k} \right)  \ Z_j^2 \ \right] + \gamma, \ j = 1, \ldots q,
 \end{equation} 
 where  $(Z_1, \ldots, Z_q) \sim Spherical_q(F_r)$. 

So, as with the SSCM, the eigenvector of the regularized M-estimators $\widehat{\Sigma}$ and $\widehat{V}$, for a given $\gamma$, are Fisher 
consistent for the eigenvectors of $\Sigma_o$, but not necessarily for the corresponding eigenvalues.  As shown in \cite{Durre:2016}, the eigenvalues 
of the SSCM functional are less dispersed than those of $\Sigma_o$, i.e.\ the SSCM is closer to being proportional to the identity matrix than 
$\Sigma_o$.  The following theorem, which uses the notation of Theorem \ref{Thrm:eigen}, shows that this is also the case in general for 
$\Sigma(F)$ and for $V(F)$. 
\begin{theorem} \label{Thrm:moreS}
Suppose $F \sim Elliptical_q(0,\Sigma_o;g)$ with $g$ being decreasing. Also, in addition to the stated assumptions on $u(s)$,
suppose $u(s) > 0$ is non-increasing. For $i, \ j = 1, \ldots, q$, if $i < j$, then for $0 < \gamma < 1$,
\[ 1 \le \frac{\lambda_i}{\lambda_j} \le \frac{\lambda_{v,i}}{\lambda_{v,j}} \le \frac{\lambda_{o,i}}{\lambda_{o,j}}, 
\mbox{ with equality if and only if  } \lambda_{o,i} = \lambda_{o,j}. \] 
Furthermore, $\lambda_i/\lambda_j$ is a decreasing function of $\gamma$, with $\lambda_i/\lambda_j = 1$ when $\gamma = 1$
and $\lambda_i/\lambda_j = \lambda_{o,i}/\lambda_{o,j}$ when $\gamma = 0$.
\end{theorem}

This theorem states that the ratio of any two eigenvalues of $\Sigma$ are closer to one than the corresponding ratio of the eigenvalues of $V$ 
which in turn are closer to one than the corresponding ratio of the eigenvalues of the population scatter matrix $\Sigma_o$, or equivalently of the
population shape matrix $\Sigma_{o,sh} = q \Sigma_o/\tr(\Sigma_o)$.
From Theorem \ref{Thrm:moreS}, it follows $\lambda_{o,i}/\lambda_{o,j} = 1 \Leftrightarrow \lambda_{i}/\lambda_{j} = 1$, and under a sequence 
of values for $\Sigma_o$, if $\lambda_{i}/\lambda_{j} \to \infty$ then $ \lambda_{o,i}/\lambda_{o,j} \to \infty$. In general, though, the values
of $\lambda_{o,i}/\lambda_{o,j}$ and  $\lambda_{i}/\lambda_{j}$ can greatly differ. This is demonstrated in Table \ref{table:cn} below.

Table \ref{table:cn}  gives the values of the condition numbers $\lambda_{o,1}/\lambda_{o,q}, \lambda_{1}/\lambda_{q}$, and 
$\lambda_{v,1}/\lambda_{v,q}$ when using the weight function $u(s) = \kappa/(s+2)$. When $\kappa = q+2$, this weight function corresponds to 
the weight function arising from the maximum likelihood estimator for  an elliptical  t-distribution on 2 degrees of freedom. The values of 
$\lambda_1, \ldots, \lambda_q$ are obtained by solving \eqref{eq:Lambda}, with the expectation evaluated by using a Monte Carlo simulation based 
on a sample size of $10^5$. The values of the condition numbers are evaluated under the multivariate normal distribution. 
In the table, $\lambda_{v,1}/\lambda_{v,q}$ slightly exceeds $\lambda_{o,1}/\lambda_{o,q}$ for the case $q = 50$, $\kappa = 53$  and 
$\gamma = 0.05$, which seems to contradict Theorem \ref{Thrm:moreS}. However, this is an artifact of the simulations since the condition 
number is a convex function and tends to be overestimated in a simulation.

\begin{table}[h]
	\centering
	\caption{Condition numbers for the regularized functionals $\Sigma(F)$ and $V(F)$ for $q = 5$ and $q = 50$
	with $F$ being multivariate normal with mean zero and covariance matrix $\Sigma_o$.
	Model 1 corresponds to the case for which the eigenvalues of $\Sigma_o$ are $(10, 1, \ldots, 1)$,
	and Model 2 corresponds to the case for which the eigenvalues of $\Sigma_o$ are uniformly spaced
	from 10 to 1.	
	The population condition number $\lambda_{o,1}/\lambda_{o,q} = 10$ for both models.	
	\label{table:cn}}
	\vspace*{8pt}
	\begin{tabular}{|c|c|ccccc|ccccc|}
	\hline
& $\kappa \downarrow$ &  &\multicolumn{3}{c}{$\lambda_{1}/\lambda_{q}$}     &   &  & \multicolumn{3}{c}{$\lambda_{v,1}/\lambda_{v,q}$}    & \\
&   $\gamma \rightarrow$ &  0.05 & 0.20  & 0.50 & 0.80  & 0.95  &  0.05 & 0.20  & 0.50 & 0.80  & 0.95 \\
\hline \hline
& 0.5 & 1.22 & 1.17 & 1.09 & 1.03 & 1.00 & 4.91 & 5.03 & 5.19 & 5.30 & 5.35 \\ 
Model 1 & 1& 1.53 & 1.40 & 1.21 & 1.07 & 1.01 & 5.37 & 5.39 & 5.39 & 5.38 & 5.37 \\ 
& 3 & 4.00 & 2.92 & 1.81 & 1.24 & 1.05 & 7.65 & 7.09 & 6.29 & 5.68 & 5.44 \\ 
q = 5 & 5 & 7.38 & 5.03 & 2.66 & 1.43 & 1.08 & 9.31 & 8.49 & 7.20 & 6.01 & 5.51 \\ 
& 8 & 9.38 & 7.43 & 4.12 & 1.79 & 1.14 & 9.94 & 9.49 & 8.27 & 6.51 & 5.62 \\ 
\hline 
& 0.5  & 1.13 & 1.11 & 1.06 & 1.02 & 1.00 & 5.91 & 5.99 & 6.12 & 6.22 & 6.26 \\ 
Model 2 & 1 & 1.31 & 1.24 & 1.13 & 1.04 & 1.01 & 6.10 & 6.15 & 6.22 & 6.26 & 6.27 \\ 
& 3 & 2.66 & 2.11 & 1.50 & 1.15 & 1.03 & 7.22 & 6.98 & 6.65 & 6.41 & 6.31 \\ 
q = 5 & 5 & 6.36 & 3.87 & 2.04 & 1.28 & 1.06 & 8.99 & 8.10 & 7.16 & 6.57 & 6.34 \\ 
& 8 &9.33 & 7.03 & 3.27 & 1.51 & 1.10 & 9.92 & 9.38 & 8.01 & 6.83 & 6.40 \\ 
\hline \hline
& 0.5 & 1.05 & 1.04 & 1.03 & 1.01 & 1.00 & 8.08 & 8.08 & 8.07 & 8.07 & 8.07 \\ 
Model 1 & 2 & 1.24 & 1.20 & 1.12 & 1.04 & 1.01 & 8.31 & 8.27 & 8.20 & 8.12 & 8.08 \\ 
& 20 & 4.16 & 3.57 & 2.49 & 1.53 & 1.12 & 9.57 & 9.46 & 9.15 & 8.63 & 8.23 \\ 
q = 50 & 50 &  7.67 & 6.43 & 4.19 & 2.14 & 1.25 & 9.94 & 9.86 & 9.60 & 9.01 & 8.38 \\ 
& 53 & 9.57 & 8.14 & 5.30 & 2.56 & 1.33 & 10.04 & 9.97 & 9.75 & 9.19 & 8.47 \\ 
\hline 
& 0.5 & 1.01 & 1.01 & 1.00 & 1.00 & 1.00 & 9.43 & 9.43 & 9.44 & 9.44 & 9.44 \\ 
Model 2 & 2 & 1.06 & 1.05 & 1.03 & 1.01 & 1.00 & 9.45 & 9.45 & 9.44 & 9.44 & 9.44 \\ 
& 20 & 1.96 & 1.74 & 1.39 & 1.13 & 1.03 & 9.60 & 9.58 & 9.52 & 9.47 & 9.45 \\ 
q = 50 & 50 & 4.74 & 3.45 & 2.03 & 1.30 & 1.06 & 9.82 & 9.74 & 9.61 & 9.51 & 9.45 \\ 
& 53 & 9.22 & 5.86 & 2.66 & 1.42 & 1.08 & 10.03 & 9.88 & 9.68 & 9.53 & 9.46 \\ 
\hline
     \end{tabular}
\end{table}

This implies, for any fixed $\gamma$, the estimators $\widehat{\Sigma}$ and $\widehat{V}$ can be heavily biased estimates of 
the population scatter matrix. This is also true for the corresponding estimates of the population shape matrix.  This bias is partially due to the 
nature of penalization, i.e.\ a trade off is made to obtain a less variable or more stable estimator, especially for smaller sample sizes, at the 
expense of generating more bias at some models.  Typically, when using  penalization methods, the bias tends to be fairly
small whenever the tuning parameter $\gamma$ is close to zero and the sample size is large. As seen in Table \ref{table:cn} , though, 
the bias can still be quite extreme in such cases, especially when considering the higher breakdown point estimates.
Due to this large universal bias over $\gamma$, any method used to choose the tuning parameter $\gamma$, such as cross-validation, would still 
result in a heavily biased estimator regardless of the sample size.

To help understand the nature of this bias, recall that when $\gamma = 0$ the M-estimating equations \eqref{eq:KL} and \eqref{eq:hyb} reduce to the 
non-penalized M-estimating equations for multivariate scatter. It is known, though, that a necessary condition for existence of a solution in the 
non-penalized case is  $\kappa \ge q$. However, the condition $(1-\gamma) \kappa < 1$ needed for Theorems \ref{Thrm:bp} and \ref{Thrm:bp2} does not
 hold whenever $\kappa \ge  q$ unless $\gamma$ is fairly large, namely $\gamma > 1-1/q$. So, for $\kappa < q$, if $\rho(s)$ is bounded below then although
$\widehat{\Sigma}$ and $\widehat{V}$ exist and are unique for any $\gamma > 0$  their limits do not exist as $\gamma \to  0$. Due to this
discontinuity, their bias does not necessarily get small as $\gamma \to 0$ for a  fixed $\kappa < q$. One could greatly reduce the bias, by choosing a weight 
function $u(s)$  for which $\kappa \ge q$, but this would result in a relatively low breakdown point. For general $u(s)$, resolving this trade-off between 
bias and breakdown is a rather formidable problem and we leave it for a future study.  Due to some properties associated with the special case 
$u(s) = \kappa/s$, though, we are able to propose a resolution for this special case in the next section. \\

 \section{The distribution-free regularized M-estimates of scatter}  \label{Sec:dfM}
In this section we consider the special case based on Tyler's weight function $u(s) = \kappa/s$. As noted in Section \ref{Sec:robust}, the distribution
of the resulting regularized estimates do not depend on from which elliptical family the data arises. Another interesting aspect of this class of
estimates is that its corresponding shape matrix is the same whether or not one uses a scale invariant version of the penalized estimate as 
defined in Section \ref{Sec:scale}. That is, if we define $\widehat{\Sigma}(\widehat{\sigma})$ to be the unique solution to 
\begin{equation} \label{eq:pt}
\Sigma=(1-\gamma) \frac{\kappa}{n} \sum_{i=1}^n \frac{x_ix_i^{\tran}}{x_i^{\tran} \Sigma^{-1} x_i}+ \gamma  \widehat{\sigma}^2I_q, 
\end{equation}
then  $\widehat{\Sigma}(\widehat{\sigma}) =   \widehat{\sigma}^2 \widehat{\Sigma}(1)$, and so
their respective shape matrices $\widehat{\Sigma}_{sh}(\widehat{\sigma})  =
q \widehat{\Sigma}(\widehat{\sigma})/\tr\{\widehat{\Sigma}(\widehat{\sigma})\}$ and $\widehat{\Sigma}_{sh}(1)  =
q\widehat{\Sigma}(1)/\tr\{\widehat{\Sigma}(1)\}$ are equal.   The shape matrix for \mbox{$\widehat{V}(\widehat{\sigma}) \equiv
 (\widehat{\Sigma}(\widehat{\sigma}) - \gamma \widehat{\sigma}^2 I_q)/(1-\gamma)$} also
 does not depend on $\widehat{\sigma}$. 
 
To obtain a compromise between bias and breakdown for the regularized M-estimates of scatter estimate, there are two tuning parameters
to consider, namely $\kappa$ and $\gamma$.  When tuning for the shape matrix using Tyler's weight function,  the problem can be
reduced to the single parameter $\beta \equiv (1-\gamma)\kappa$. This follows, since as observed in \cite{Ollila-Tyler:2014}, for $0 < \beta < q$, 
if we let $\widehat{\Sigma}_{\beta,\gamma}$ denote the unique solution to 
\begin{equation} \label{eq:beta}
\Sigma= \frac{\beta}{n} \sum_{i=1}^n \frac{x_ix_i^{\tran}}{x_i^{\tran} \Sigma^{-1} x_i}+ \gamma  I_q, 
\end{equation}
then for any two values $\gamma_1$ and $\gamma_2$, $\widehat{\Sigma}_{\beta,\gamma_1} = \frac{\gamma_1}{\gamma_2} \widehat{\Sigma}_{\beta,\gamma_2}$. 
Also, for $\widehat{V}_{\beta,\gamma} = (\widehat{\Sigma}_{\beta,\gamma} - \gamma I_q)/(1-\gamma)$, we have the relationship 
$\widehat{V}_{\beta,\gamma_1} = \frac{\gamma_1 (1-\gamma_2)}{\gamma_2 (1-\gamma_1)} \widehat{V}_{\beta,\gamma_2}$. Consequently, the shape matrices 
$\widehat{\Sigma}_{sh,\beta} = q \widehat{\Sigma}_{\beta,\gamma}/\tr\{\widehat{\Sigma}_{\beta,\gamma}\}$  and 
$\widehat{V}_{sh,\beta} =q \widehat{V}_{\beta,\gamma}/\tr\{\widehat{V}_{\beta,\gamma}\}$ do not depend on the value of $0 < \gamma <1$.  
 
 As noted in Section \ref{Sec:robust}, some conditions on the sample are needed to ensure \eqref{eq:beta} admits a unique solution. These conditions,
 given by Condition A in \cite{Ollila-Tyler:2014}, place some restrictions on how many data points can lie within subspaces. Specifically, the inequaltiy 
 $\# \{x_i \in \mV\}/n < dim(\mV)/\beta$ must hold for any subspace $\mV$ of $\R^q, 1 \le dim(\mV) < q$.  This 
 condition always holds when $\beta < 1$. When the data are in \emph{general position}, i.e.\ when every subset of $r \le q$ data points spans 
 a subspace of dimension $r$, the condition holds for any $0 < \beta < q$. When randomly sampling from a continuous distribution in $\R^q$, the data
 are in general position with probability one.
 
 Using the results on existence, we can extend the breakdown point results given in Section \ref{Sec:BP}. The proof of the following theorem is given
 in the appendix.
\begin{theorem} \label{Thrm:bdbeta} 
Suppose $n > \beta$ for $\widehat{\Sigma}_{\beta,\gamma}(X)$ and $n \ge q$ for $\widehat{V}_{\beta,\gamma}(X)$. If the sample $X=\{x_1, \cdots, x_n\}$ 
is in general position, then
\begin{itemize}
\item[i)] for $0 \le \beta < 1, \  \epsilon^*(X;\widehat{\Sigma}_{\beta,\gamma}) = \epsilon^*(X;\widehat{V}_{\beta,\gamma}) =  1$, and
\item[ii)] for $1 \le \beta < q, \  \epsilon^*(X;\widehat{\Sigma}_{\beta,\gamma}) = \epsilon^*(X;\widehat{V}_{\beta,\gamma}) 
\ge  (n-\beta)/\{(n-1)\beta\} \to 1/\beta$ as $n \to \infty$.
\end{itemize}
\end{theorem} 
The above breakdown points hold for a fixed $\beta$, but do not necessarily apply when $\beta$ is determined by the data. As noted in
the forthcoming section, it is possible for the data driven choice of $\beta$ to be close to $q$ and the resulting tuned scatter matrix to have
a high breakdown point.

 \begin{remark} \label{rem:scat}
 As noted above, neither $\widehat{\Sigma}_{sh,\beta}(X)$ nor $\widehat{V}_{sh,\beta}(X)$ depend on the value of $\gamma$. 
 The value of $\gamma$ only affects the scale of the scatter matrix. Rather than let $\gamma$ determine the scale, an alternative to
  $\widehat{\Sigma}_{\beta,\gamma}(X)$ is to define $\widehat{\Sigma}_{sc,\beta}(X) = 
 \widehat{\sigma}_\beta^2(X) \widehat{\Sigma}_{sh,\beta}(X)$, where \mbox{$\widehat{\sigma}_\beta^2(X) = 
 median\{ x_i^ \tran \widehat{\Sigma}_{sh,\beta}(X)^{-1} x_i; i = 1, \ldots, n\}/q$.} The scatter statistic $\widehat{V}_{sc,\beta}(X)$
 can be defined analogously.
Here both  $\widehat{\Sigma}_{sh,\beta}(X)$ and  $\widehat{V}_{sh,\beta}(X)$ are scale invariant, whereas 
$\widehat{\Sigma}_{sc,\beta}(X)$ and $\widehat{V}_{sc,\beta}(X)$ are scale equivariant, i.e. 
$\widehat{\Sigma}_{sc,\beta}(cX) = c^2 \widehat{\Sigma}_{sc,\beta}(X)$ for any $c \ne 0$. 
The values of $x_i = 0$ contribute to the value of $\widehat{\sigma}_\beta^2(X)$, whereas they do not contribute 
to the value of the shape statistics $\widehat{\Sigma}_{sh,\beta}(X)$. Although the shape statistics are distribution free under random samples from elliptical 
distributions, the distribution of  $\widehat{\sigma}_\beta^2(X)$ would be dependent on the specific elliptical distribution.  
\end{remark} 

\subsection{Choosing the tuning parameter \boldmath{$\beta$}}
Oracle methods for choosing tuning parameters for variants of the regularized distribution-free M-estimators of scatter have been proposed in the literature, 
see e.g.  \cite{Chen-Wiesel-Hero:2011, Ollila-Tyler:2014}. It is not clear, though, how severely the final tuned estimators are affected by outliers. 
In this section, we focus on the shape matrix, which is well defined within the semi-parametric 
class of elliptical distributions, and consider a more data dependent approach to tuning, namely cross-validation over $0 \le \beta \le q$. In deciding on a value of 
$\beta$, the resulting estimate of shape $\widehat{\Sigma}_{sh,\beta}$ or $\widehat{V}_{sh,\beta}$ should be a good fit to the data. 
We consider as a cross-validation criterion the negative log-likelihood associated with the angular central Gaussian distribution \cite{Tyler:1987b}, which for a
subsample  $\{x_{*1}, \ldots, x_{*r}\} \subset \{x_{1}, \ldots, x_{n}\}$ is given by
 \begin{equation} \label{eq:acg}
 CV(\Sigma_*;x_{*1}, \ldots, x_{*r}) = \frac{q}{r} \sum_{i=1}^r \log \left(\frac{x_{*i}^{\tran} \Sigma_*^{-1} x_{*i}}{x_{*i}^{\tran} x_{*i}} \right) + \log\{ \det(\Sigma_*)\},
 \end{equation}
 where $\Sigma_*$ is an estimate of $\Sigma$ or $V$  based on a subsample of size $n-r$ not containing 
 $\{x_{*1}, \ldots, x_{*r}\}$.  Observe that  $CV(\Sigma_*;x_{*1}, \ldots, x_{*r}) = CV(\lambda \Sigma_*;x_{*1}, \ldots, x_{*r})$ for any $\lambda>0$, and 
 so the function $CV$ only depend on $\Sigma_*$ through its shape matrix $\Sigma_{*,sh} = \Sigma_*/\tr(\Sigma_*)$. Also, the $CV$ value is invariant under 
 multiplication of the data points by constants, i.e. under the transformation $\{x_{*1}, \cdots, x_{r}\} \to \{c_1 x_{*1}, \cdots, c_r x_{*r}\}$ for any 
 set of constants $c_1, \ldots, c_r$.
 
\begin{figure}[!htbp] 
\vspace*{-.5in}
\begin{center} \large \textbf{CROSS VALIDATION METHODS APPLIED TO \boldmath{$\widehat{\Sigma}_{sh,\beta}$}} \\[16pt] \end{center}
\hspace*{-1cm}\scalebox{1}{\includegraphics{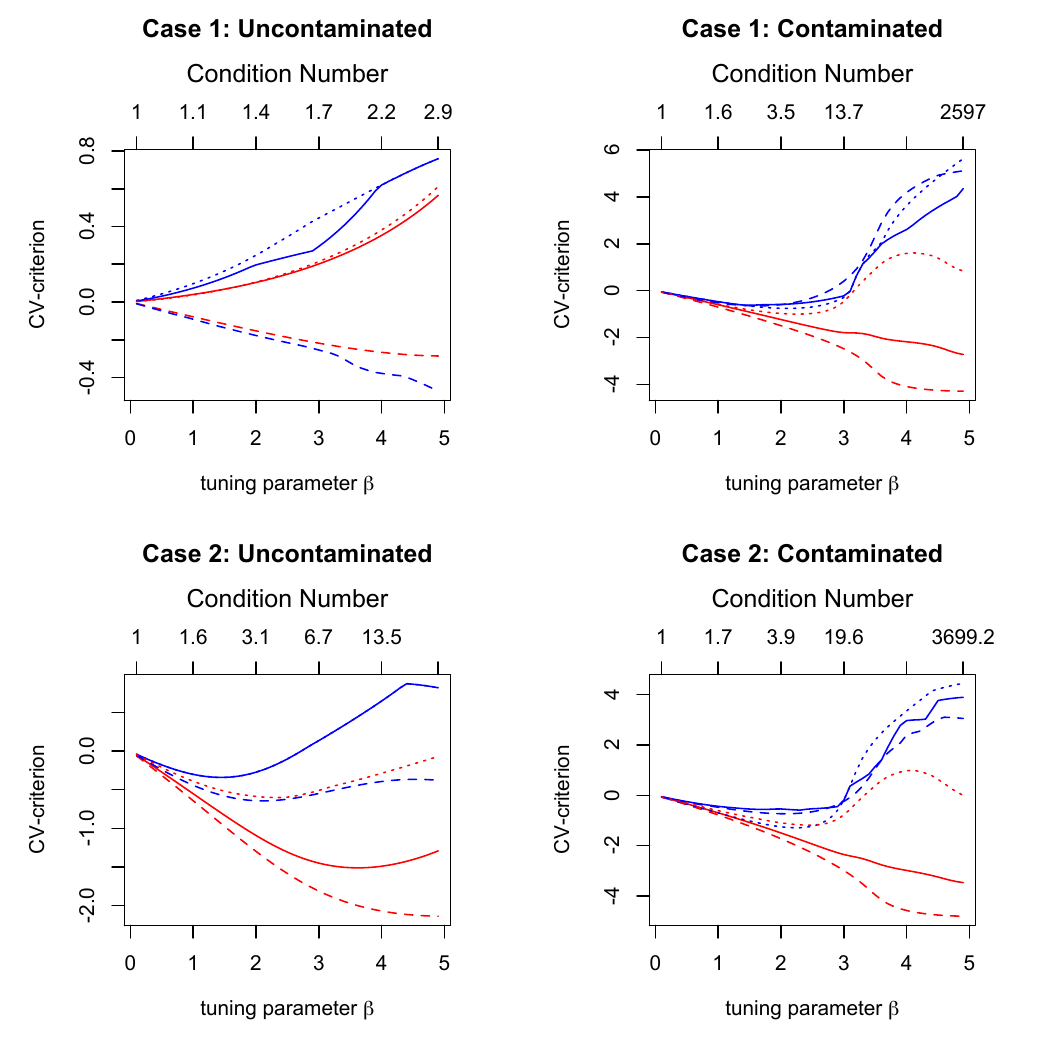}}
\caption{Different validation methods for $\widehat{\Sigma}_{\beta}$: \textbf{cvmean} - solid red , \textbf{cvmed}  - solid blue,  \textbf{cvmeanmed}  - dotted red, 
\textbf{cvmedmed} -  dotted blue, \textbf{acvr} -  dashed blue, \textbf{acv} - dashed red. \label{fig:cvplot}}
\end{figure}

\begin{figure}[!htbp] 
\vspace*{-.5in}
\begin{center} \large \textbf{CROSS VALIDATION METHODS APPLIED TO \boldmath{${\widehat{V}_{sh,\beta}}$}} \\[16pt] \end{center}
\hspace*{-1cm}\scalebox{1}[1]{\includegraphics{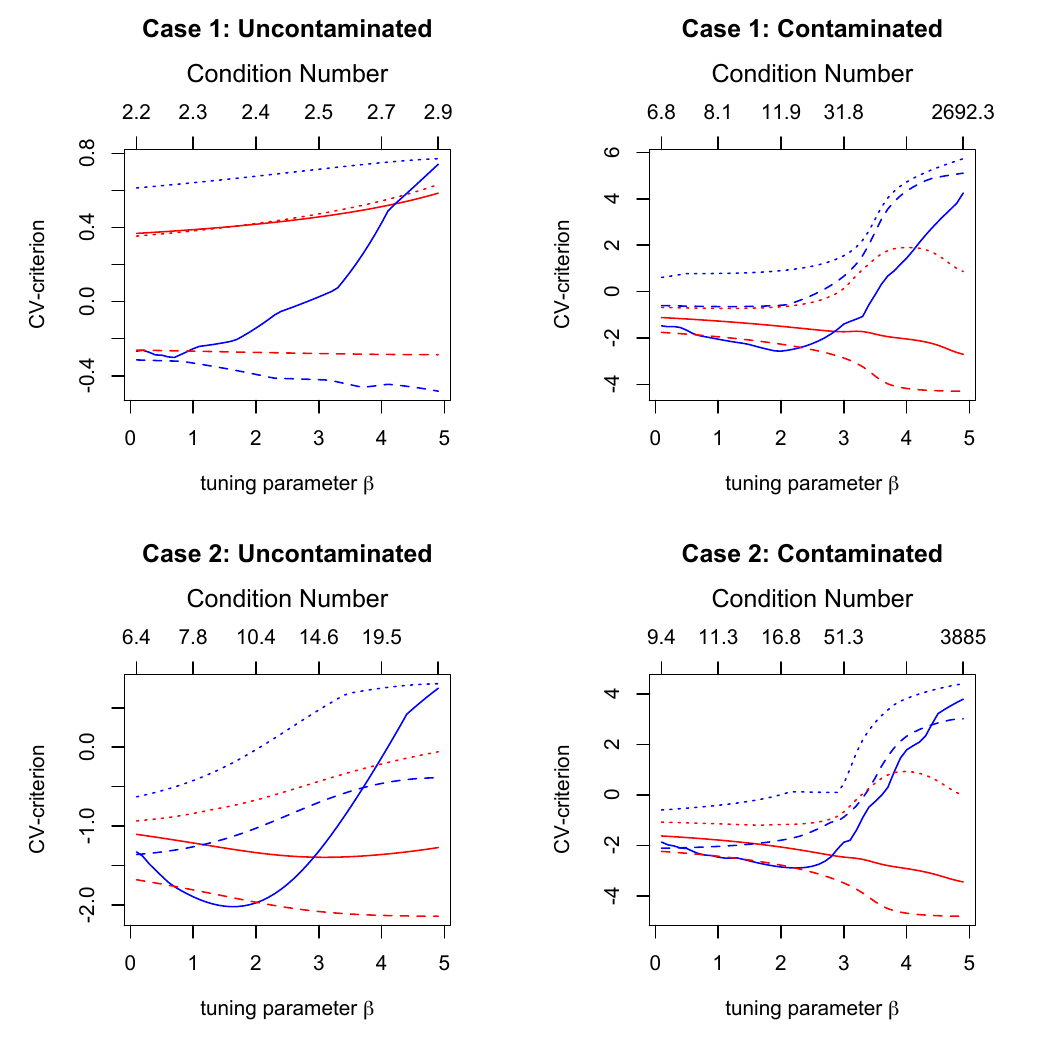}}
\caption{Different validation methods for $\widehat{V}_{\beta}$: \textbf{cvmean} - solid red , \textbf{cvmed}  - solid blue,  \textbf{cvmeanmed}  - dotted red, 
\textbf{cvmedmed} -  dotted blue, \textbf{acgmed} -  dashed blue, \textbf{acgmean} - dashed red. \label{fig:cvplotV}}
\end{figure}

 Figures \ref{fig:cvplot} and \ref{fig:cvplotV} illustrates the use of 5-fold cross validation over $\beta$ for $\widehat{\Sigma}_{sh,\beta}$ and 
 $\widehat{V}_{sh,\beta}$ respectively, with the grid for $\beta$ being in increments of $0.1$. The top axes gives the condition number $(cn)$ for the given 
 value of $\beta$.  Cases 1 and 2 correspond to random samples 
 of size $n =  35$ from a $q=5$ dimensional multivariate  normal distribution with mean zero and respective covariance matrices $I_5$ and $\diag(10, 1, 1, 1, 1)$, 
 and so the corresponding condition numbers are 1 and 10.  The contaminated cases in the figures are generated by randomly replacing $10$ of the good 
 observations with the symmetric contaminating points $\pm \{y_1, \ldots, y_5\}$, with $\{y_1, \ldots, y_5\}$ being a random sample from a $q = 5$ 
 dimensional multivariate normal distribution with mean $(5, \ldots, 5)$ and covariance matrix $\diag(0.01, \ldots, 0.01)$. The proportion of contamination is thus
 $10/35$ or $28.6\%$. The cross validation curves in Figures \ref{fig:cvplot} and \ref{fig:cvplotV} given by  
 the average of \eqref{eq:acg} over the five validation subsamples is denoted as the \textbf{cvmean} criterion and is represented by the 
 \red{solid red} lines in the figures. For the uncontaminated case 1, the minimum is obtain at $\beta = 0$ for both
 $\widehat{\Sigma}_{sh,\beta}$ and $\widehat{V}_{sh,\beta}$, i.e.\ the perfect estimate $I_5 \ (cn = 1)$ and the SSCM $(cn = 6.50)$ respectively.
 For  the uncontaminated case 2, the minimum for $\widehat{\Sigma}_{sh,\beta}$ is obtain at $\beta = 3.6 \ (cn= 10.37)$  and for $\widehat{V}_{sh,\beta}$
 at $\beta = 3.1 \ (cn=15.12)$.  For these two cases, the estimates seems reasonable, with the estimate based on $\widehat{\Sigma}$ being better than that
 based on $\widehat{V}$. For the contaminated cases, though, the minimum is always obtained at $\beta =5$ with  $cn = \infty$, i.e.\ the estimates break down. 
 
Besides providing a good fit to the data when the data is not contaminated,  another consideration when deciding on a value of $\beta$ is that the resulting estimate 
of shape should have good breakdown properties, say does not breakdown under $\epsilon < 50 \%$ contamination. This implies that the condition number should 
be bounded above under $\epsilon-$contamination. From Theorem  \ref{Thrm:bdbeta}, a breakdown point of at least $50 \%$ can be obtained if we restrict cross 
validation to $\beta \le 2$. However, this may not necessarily yield an estimate which is a good fit to the data when little or no contamination is present, especially 
for large $q$. We therefore propose the following alternative approach. 
 
When the data is contaminated, the cross validation criterion measures how well a shape matrix fits the contaminated data set, i.e. the mixture 
of the good and bad data points.  As noted in \cite{Ronchetti:1997}, outliers may place undue influence on the chosen model or in
our case the chosen value of $\beta$.  To place more weight on how well the good data are fitted,
we propose replacing the average in \ref{eq:acg} with the median to obtain the more robust criterion
 \begin{equation} \label{eq:med-acg}
 CV_R(\Sigma_*;x_{*1}, \ldots, x_{*r}) =  
 median\left\{ q  \log \left(\frac{x_{*i}^{\tran} \Sigma_*^{-1} x_{*i}}{x_{*i}^{\tran} x_{*i}} \right) + \log\{ \det(\Sigma_*)\}; 
 \  i =1, \ldots, r \right\}.
 \end{equation}
 Note that other measures of center besides the average and the median could also be considered in defining the cross validation criteria.   
 A similar concept has been proposed in the context of ridge regression in \cite{Jung:1991}.
 Analogous to the criterion $CV$, this $CV_R$ criterion also depends only on the shape matrix $\Sigma_{*,sh} = \Sigma_*/\tr(\Sigma_*)$, and is 
 invariant under multiplication of the data points by constants.  
 
 A motivation for considering \eqref{eq:med-acg} is to observe that the value of $\Sigma > 0$ which minimizes \eqref{eq:acg} for the full data set, 
 i.e.\ $\widehat{\Sigma}_T = \arg\inf_{\Sigma > 0}  CV(\Sigma; x_1, \ldots, x_n)$, corresponds to Tyler's distribution-free M-estimate of scatter which 
 has a breakdown point of  $1/q$. However, as shown in Theorem  \ref{thm:sbd} below, the breakdown point of 
 $\widehat{\Sigma}_R = \arg\inf_{\Sigma > 0}  CV_R(\Sigma; x_1, \ldots, x_n)$  goes to $1/2$ as $n \to \infty$. We wish to emphasize
 that the estimate $\widehat{\Sigma}_R$ introduced here corresponds to a new affine equivariant high breakdown point scatter statistic, see Remark
 \ref{rem:affine}.  Although $\widehat{\Sigma}_T$ is 
 known to be readily computed via a simple reweighing  algorithm, the statistic $\widehat{\Sigma}_R$, as is common with all affine equivariant high  breakdown 
 point scatter statistics, appears to be computationally intensive. However, we do not need to compute $\widehat{\Sigma}_R$ for our purposes. Rather, we only
 need to minimize $CV_R(\Sigma_*;x_{*1}, \ldots, x_{*r})$ over the candidate values of $\Sigma_*$.  
 
In K-fold cross validation, some subsamples obtained from the partitioning may contain over 50\% bad data points when the data is highly 
contaminated.  So, rather than taking the average of $CV_R$ over the K validation subsamples, we propose either replacing the average with the median, and 
i.e.\  taking the median of the medians, or to simply use
 \begin{equation} \label{eq:medall}
 median \left\{ q \log \left(\frac{x_{*i}^{\tran} \Sigma_*^{-1} x_{*i}}{x_{*i}^{\tran} x_{*i}} \right)  + \log\{ \det(\Sigma_*); \ i = 1, \ldots n \right\},
 \end{equation}
 i.e.\ the median over all $n$ such values arising in \eqref{eq:med-acg} from the K validation subsamples. These different cross validation approaches are also 
 illustrated in Figures \ref{fig:cvplot} and \ref{fig:cvplotV}.  The cross validation curves obtained using \eqref{eq:medall} is denoted the \textbf{cvmed} criterion and is 
 represented by the \blue{solid blue} lines in the figures, whereas those associates with taking the mean and median of \eqref{eq:med-acg} over the five 
 validation subsamples are denoted the \textbf{cvmedmean} and the \textbf{cvmedmed} criterion respectively, and are represented by the 
 \red{dotted red}  and the \blue{dotted blue} lines in the figures. The \blue{dotted blue} line coincides with the \blue{solid blue} line in Figure \ref{fig:cvplot} in
 the uncontaminated case 2.  
 
 For the uncontaminated case 1, the minimum is again obtain at $\beta = 0$ for any of the above criterion, with the exception of \textbf{cvmed} for 
 $\widehat{V}_{sh,\beta}$ for which the minimum is obtained at $\beta = 0.7 \ (cn = 2.28)$.  So, all of the estimates based on the four different cross validation
 criterion give comparable results for this case. For  the uncontaminated case 2, the minima for 
 $\widehat{\Sigma}_{sh,\beta}$ are obtained at $\beta = 1.4 \ (cn = 2.09)$, $\beta = 2.3 \ (cn = 3.93)$  and $\beta = 1.4 \ (cn =  2.09)$  for \textbf{cvmed},   
  \textbf{cvmedmean} and \textbf{cvmedmed} respectively, and the minima for $\widehat{V}_{sh,\beta}$ are respectively obtained at 
  $\beta = 1.6 \ (cn = 9.20)$, $\beta = 0 \ (cn = 6.50)$  and $\beta = 0 \ (cn =  6.50)$.  For this case, the estimates based on the robust cross validation criterion 
  are comparable to the estimates based on the \textbf{cvmean}, with the exception that the error in the condition number for $\widehat{\Sigma}_{sh,\beta}$ 
  is smallest for the criterion \textbf{cvmean}, whereas for $\widehat{V}_{sh,\beta}$ it is smallest for the criterion \textbf{cvmed}.  Overall, the robust cross validation 
  criterion appears to be minimized at smaller values of $\beta$  than \textbf{cvmean}, i.e. they tend to more heavily shrink the estimate towards proportionality
  to the identity. 
  
  For the contaminated cases, though, we see that the estimates based on the robust cross validation criterion do not breakdown. Also, the errors in the the estimated condition numbers are not too large even under close to $30\%$ contamination.  In particular, the minima for $\widehat{\Sigma}_{sh,\beta}$ based on \textbf{cvmed},   
 \textbf{cvmedmean} and \textbf{cvmedmed} respectively are $\beta = 1.5 (cn = 2.3)$, $\beta = 2.3 (cn =  4.78)$ and $\beta = 2.1 (cn =  3.88)$ 
for contaminated case 1 and $\beta = 2.3 \ (cn = 5.61)$, $\beta = 2.4 \ (cn =  6.42)$ and $\beta = 2.2 \ (cn =  4.95)$  for contaminated case 2. The corresponding
values for $\widehat{V}_{sh,\beta}$ are $\beta = 2.0 \ (cn = 11.97)$, $\beta = 1.1 \ (cn =  8.39)$ and $\beta = 0 \ (cn =  6.80)$  for contaminated case 1 and 
$\beta = 2.2 \ (cn = 19.23)$, $\beta = 1.6 \ (cn =  13.86)$ and $\beta = 0 \ (cn =  9.49)$  for contaminated case 2. One can note that for the contaminated cases,
 the curves associated with \textbf{cvmeanmed} tend to curve downward for larger values of $\beta$,  although not enough to cause breakdown since  the 
 minima are not obtained at the upper extreme.  Also, the curves for \textbf{cvmed} tend to be less smooth than the other curves.  Based upon these limited
 results, we believe \textbf{cvmedmed} to be the best criterion to use for cross validation. It is also interesting to note that the robust cross validation curves for the
 contaminated cases tend to sharply increase near $\beta = 3$. This agrees with Theorem \ref{Thrm:bdbeta} which states that the breakdown point at $\beta = 3$ is
 approximately $33.3\%$, whereas the proportion of contamination is  $28.6\%$, and suggests the robust cross-validation curves may help ascertain the 
 approximate percentage of the data which may be contaminated.
 
Another method for choosing the tuning parameter $\beta$ without using cross validation is to simply minimize the robust cross-validation criterion 
\eqref{eq:med-acg} over $\widehat{\Sigma}_\beta$ without using subsamples. That is, define
 \begin{equation} \label{eq:betahat}
 \widetilde{\beta} =  \arg\inf { }_{ 0 \le \beta < q } \ CV_R(\widehat{\Sigma}_\beta; x_1, \ldots, x_n),
  \end{equation}
with the resulting tuned penalized shape matrix then being $\widehat{\Sigma}_{sh,\widetilde{\beta}}$, or $\widehat{V}_{sh,\widetilde{\beta}}$ when 
$\widehat{\Sigma}_\beta$ is replaced by $\widehat{V}_\beta$ in \eqref{eq:betahat}.  Choosing the tuning parameter $\beta$ in this manner requires 
less computations than cross validation methods.  Also, as noted in Theorem \ref{thm:sbd} below, the breakdown point of the resulting tuned shape matrix 
is approximately $1/2$ for large $n$.  
 
 \begin{theorem} \label{thm:sbd}
If $X = \{x_i \in \R^q; i  = 1,  \ldots, n \}$ is in general position, then the finite sample contamination breakdown point for $\widehat{\Sigma}_{sh}$, where
 $\widehat{\Sigma}_{sh}$ is either  $\widehat{\Sigma}_{sh,R}$, $\widehat{\Sigma}_{sh,\widetilde{\beta}}$ or $\widehat{V}_{sh,\widetilde{\beta}}$, satisfies 
\[ \epsilon^*(X; \widehat{\Sigma}_{sh})  \ge \frac{n-2q+2}{2n-2(q-1)} \to \frac{1}{2} \ \mbox{as} \ n \to \infty. \]
\end{theorem}
\noindent
Obtaining exact results for the breakdown point for the shape estimates based on $\beta^*$, the value of $\beta$ obtain via cross validation, is a more challenging 
problem due to the randomness of $\beta^*$ arising from the simulated partitioning of the data. We leave this as an open problem, although based on 
Figures \ref{fig:cvplot} and \ref{fig:cvplotV} we conjecture that the breakdown points will be near $1/2$ when using the median based cross validation methods, 

The behavior of the criterion $CV_R(\widehat{\Sigma}_\beta; x_1, \ldots, x_n)$, denoted as \textbf{acgmed} and represented by a \blue{dashed blue}
curve, is also illustrated in Figures \ref{fig:cvplot} and \ref{fig:cvplotV}. For completeness we also include in the figures the criterion 
 $CV(\widehat{\Sigma}_\beta; x_1, \ldots, x_n)$ which is denoted by \textbf{acgmean} and represented by the \red{dashed red} curve. Note that the
  \textbf{acgmean} curve is always minimized at $\beta = 5$ since as previously noted $\widehat{\Sigma}_q \widehat{\Sigma}_T$ which 
  minimizes $CV(\widehat{\Sigma}; x_1, \ldots, x_n)$ over all $\Sigma > 0$. The \textbf{acgmed} curve is also minimized when $\beta = 5 \ (cn = 2.97)$ for both
  $\widehat{\Sigma}_{sh,\beta}$ and $\widehat{V}_{sh,\beta}$ for the uncontaminated case 1. For the other cases, \textbf{acgmed} curve tends
  to have properties similar to the \textbf{cvmedmed} curve, which suggests its use may provide a computationally simpler viable estimate, especially for
  large sample sizes and/or high dimensions.  In particular, for the uncontaminated case 2, the \textbf{acgmed} curves in Figures \ref{fig:cvplot} and \ref{fig:cvplotV}
  are minimized at $\beta = 2.1 \ (cn = 3.39)$ and $\beta = 0 \ (cn = 6.50)$ respectively. For the contaminated cases, the curves for case 1 are respectively minimized at
   $\beta = 1.6 \ (cn = 2.53)$ and $\beta = 1.2 \ (cn = 8.64)$ and those for case 2 are minimized at  $\beta = 2.1 \ (cn = 4.41)$ and $\beta = 0 \ (cn = 9.49)$. 
  
\begin{remark} \label{rem:affine}
The shape statistic $\widehat{\Sigma}_{sh,R}(X)$ is equivariant under linear transformations of the data in the sense that if $A$ is 
non-singular, then $\widehat{\Sigma}_{sh,R}( AX) \propto  A\widehat{\Sigma}_{sh,R}(X) A^\tran$. 
If we define the scatter statistic $\widehat{\Sigma}_{sc,R}(X) = \widehat{\sigma}_R^2(X) \widehat{\Sigma}_{sh,R}(X)$, 
where  $\widehat{\sigma}_R^2(X) = median\{ x_i^\tran \widehat{\Sigma}_{sh,R}(X)^{-1} x_i; i = 1, \ldots, n\}/q$, then it 
satisfies $\widehat{\Sigma}_{sc,R}(AX) = A \widehat{\Sigma}_{sc,R}(X) A^ \tran$.   The breakdown point for $\widehat{ \Sigma}_{sc,R}( X)$
is the same as the breakdown point for $\widehat{ \Sigma}_{sh,R}( X)$ given in Theorem \ref{thm:sbd}. 

Using a technique introduced  in Kent and Tyler (1991), one can generate an affine equivariant multivariate location and scatter statistic by 
appending a one to 
each observation in $X$, that it define $X_o = \{ x_{o,i} \in \R^{q+1}; i  = 1,  \ldots, n \}$ where $ x_{o,i}^ \tran = ( x_{i}^ \tran, 1)$,  and
then partition
\[ \widehat{ \Sigma}_{sc,R}( X_o) = \left[ \begin{array}{cc}  
\widehat{ \Sigma}  + \alpha \widehat{ \mu} \widehat{ \mu}^ \tran & \alpha \widehat{ \mu} \\
\alpha \widehat{ \mu}^ \tran & \alpha
\end{array} \right],
\]
The resulting statistics  $(\widehat{ \mu}, \widehat{ \Sigma})$ are high breakdown point affine equivariant multivariate location and scatter statistics. 
\end{remark}

\section{Concluding remarks and future directions} \label{Sec:conclude}
A motivation for considering the regularized M-estimators of scatter is to observe that the SSCM does
not take into account the shape of the data cloud when down-weighing observations, but rather down-weights  based on their Euclidean distances from the 
center.  On the other hand,  the non-penalized M-estimators down-weights observations based on their adaptive Mahalanobis distances. However, in addition to  
their relatively low breakdown points in higher dimension, these M-estimators may be ill-conditioned when the sample size is of the same order as the dimension 
of the data. The regularized M-estimators of scatter can be seen as a compromise between these approaches to down-weighing observations. 
They take into account the shape of the data cloud, but adjust the Mahalanobis contours towards a more spherical shape. This helps prevent problems 
which may arise from inverting an ill-conditioned scatter matrix. Overall our proposed method has ease of computation and the stability of an M-estimate while 
being a high breakdown method resistant to outliers.

Although the emphasis on tuning in this paper is on the distribution-free regularized M-estimators of shape, the cross-validation \eqref{eq:acg} 
and the robust cross-validation \eqref{eq:med-acg} criteria for shape can be used for regularized M-estimators in general. In this more general case, cross validation 
needs to be done over the parameters $\kappa > 0$ and $0 \le \gamma \le 1$. Here, the parameter $\kappa$ can be defined within a class of weight functions
$u(s) = \kappa u_o(s)$, with $u_o(s)$ being a fixed function with $sup_{s > 0} \psi_o(s) = 1$ and hence $sup_{s > 0} \psi(s) = \kappa$. The idea is to still focus on 
the shape of the scatter functional, which is well defined within the semi-parametric class of elliptical distributions. Rather than use the scale associated with 
$\widehat{\Sigma}_{\kappa,\gamma}$ or $\widehat{V}_{\kappa,\gamma}$, we again propose defining scale as in Remark \ref{rem:scat}.  For example, given 
the estimate for shape $\widehat{\Sigma}_{sh,\kappa,\gamma} = q \ \widehat{\Sigma}_{\kappa,\gamma}/\tr(\widehat{\Sigma}_{\kappa,\gamma})$, define the
regularized estimate of scatter as $\widehat{\Sigma}_{sc,\kappa,\gamma} = \widehat{\sigma}^2 \widehat{\Sigma}_{sh,\kappa,\gamma}$ where 
$\widehat{\sigma}^2 = C * med(x'\widehat{\Sigma}_{sh,\kappa,\gamma}x)$, with the constant $C$ is customarily chosen so that $\widehat{\Sigma}$ is consistent 
at the multivariate normal model. Also, we believe the breakdown point results given Theorem \ref{Thrm:bdbeta} can be extended to general weight functions 
$u(s)$ and in particular we conjecture for this more general setting that the breakdown point is approximately $\min(1/\{(1-\gamma)\kappa\},1)$. 

The high breakdown point estimator  $\widehat{\Sigma}_R = \arg\inf_{\Sigma > 0}  CV_R(\Sigma; x_1, \ldots, x_n)$ discussed in 
Remark \ref{rem:affine} is a special case of a maximum trimmed likelihood estimator (MTLE), or more specifically a maximum median likelihood estimator (MMLE), as
defined in \cite{Hadi-Luceno:1997}.  In particular, the estimator  $\widehat{\Sigma}_R$  is the MMLE under the angular central Gaussian distribution, which can be 
viewed as being at the extreme end of elliptical distributions. A general class of cross-validation criteria can be constructed using the objective function associated
with an MTLE or MMLE for other elliptical distributions, and their properties would be worth exploring in future research. It is worth noting that it is shown in
\cite{Hadi-Luceno:1997} the MMLE under the multivariate normal model is the MVE. As emphasize in our paper, using the angular central Gaussian model has the 
advantage of depending only on the well-defined shape matrix of the elliptical distribution, as well as being distribution-free over the class of elliptical distributions.

 Finally, we note that a an R-package is being developed to implement the methods proposed in this paper, and that a large simulation study for the tuned regularized 
 M-estimates of scatter is on-going and will be reported elsewhere.

\section*{Acknowledgements}
Research for David E.\ Tyler's and Mengxi Yi was supported in part by National Science Foundation Grants DMS-1407751
and DMS-1812198. Mengxi Yi’s research was also supported in part through Austrian Science Fund (FWF) under grant P31881-N32, the
National Natural Science Foundation of China (No.\ 12101119) and the Fundamental Research Funds for the Central Universities
(310422113). \\

\section*{Appendix: Proofs}\label{Sec:Proofs}

 \subsection*{\small \emph{Proof of Theorem \ref{Thrm:bp}}}
It is immediate from \eqref{eq:OT} that $\widehat{\Sigma}_1 \ge \eta I_q$. To obtain an upper bound, express \eqref{eq:OT} as
\begin{align*}
\Sigma &= \n\psi\left(x_i^{\tran}\Sigma^{-1}x_i\right)\frac{x_ix_i^{\tran}}{x_i^{\tran}\Sigma^{-1}x_i}+\eta I_q \\
&\le \frac{\kappa \lambda_1}{n} \sum_{i=1}^n \frac{x_ix_i^{\tran}}{x_i^{\tran}x_i}+\eta I_q \le (\kappa \lambda_1+\eta) I_q,
\end{align*}
where $\lambda_1$ denotes the largest eigenvalue of $\Sigma$.  This then implies $\lambda_1 \le \kappa \lambda_1+\eta$ or
$\lambda_1 \le \eta/(1-\kappa)$. Thus,  $\widehat{\Sigma}_1 \le \frac{\eta}{1-\kappa} I_q$. The results for $\widehat{\Sigma}_2$ follow
by relating equations \eqref{eq:OT} and \eqref{eq:KL}.

 \subsection*{\small \emph{Proof of Theorem \ref{Thrm:bp2}}}
Consider the contaminated data set \mbox{$Z = X \cup Y$}, where $Y = \{y_1, \ldots, y_m\}$. For this data set, denote 
the solution to \eqref{eq:hyb2} by $\Vhh$, and let $\lambda_q$ be the smallest eigenvalue of $\Vhh$. Since 
\[z^{\tran}\{\Vhh+\eta I_q\}^{-1}z \le z^{\tran}z/(\lambda_q+ \eta),\] 
it follows from the non-increasing property of the weight function $u(s)$ that 
\[ u(z^{\tran}\{\Vhh+\eta I_q\}^{-1}z) \ge u\left(z^{\tran}z/(\lambda_q+ \eta)\right).\]
Thus, from \eqref{eq:hyb2} we obtain
\begin{align*}
  \Vhh  & \ge \frac{1}{n+m} \sum_{i=1}^{n+m} u\left(\frac{z_i^{\tran}z_i}{\lambda_q + \eta}\right)z_iz_i^{\tran}\\ 
    & = \frac{\lambda_q+ \eta}{n+m}\sum_{i=1}^{n+m}\psi\left(\frac{z_i^{\tran}z_i}{\lambda_q+\eta}\right)\frac{z_iz_i^{\tran}}{z_i^{\tran}z_i} \\
        & \ge \frac{\lambda_q+ \eta}{n+m}\sum_{i=1}^{n}\psi\left(\frac{x_i^{\tran}x_i}{\lambda_q+\eta}\right)\frac{x_ix_i^{\tran}}{x_i^{\tran}x_i} \end{align*}
Therefore, $\Vhh$ is bounded below over all possible contaminations $Y$ as long as the good data $X$ spans $\R^q$. By Theorem \ref{Thrm:bp}
it follows that $\Vhh$ is also bounded above over all $Y$. Hence it cannot break down. The proof for $\widehat{V}_2$ is analogous.

\subsection*{\small \emph{Proof of Theorem \ref{Thrm:eigen}}}
\textbf{Part a:} We first show that the matrices $\Sigma \equiv \Sigma(F)$ and $\Sigma_o$ are jointly diagonalizable.  That is, they possess the same eigenvectors.
To prove this, let $Z = \Sigma^{-1/2}_o X \sim Spherical_q$. Pre- and post-multiplying \eqref{eq:func} by $\Sigma^{-1/2}_o$ gives:
\begin{equation} \label{eq:T2}
\Gamma \equiv \Sigma^{-1/2}_o\Sigma\Sigma^{-1/2}_o = (1-\gamma) E[u(Z^\tran \Gamma^{-1} Z) Z Z^\tran] + \gamma \Sigma^{-1}_o. \end{equation}
Denote the spectral value decomposition of $\Gamma = P \Delta P^\tran$. Since  $Z \sim P^\tran Z$, pre- and post-multiplying \eqref{eq:T2} by  $P^\tran$ and $P$ respectively gives:
\begin{equation} \label{eq:T3} 
\Delta =  (1-\gamma) E[u(Z^\tran \Delta^{-1} Z) Z Z^\tran] + \gamma \Lambda^{-1}_o, 
\end{equation}
where $\Lambda_o \equiv P^\tran \Sigma_o P$. Since the distribution of $Z = (Z_1, \ldots, Z_q)^\tran$ is invariant under sign changes of its components, mapping
 $(Z_i, Z_j) \to  (- Z_i, Z_j)$ implies
\[ E[u( Z^\tran \Delta^{-1} Z) Z_i Z_j] = - E[u( Z^\tran \Delta^{-1} Z) Z_i Z_j] = 0. \]
Hence, the matrix $E[u(Z^\tran \Delta^{-1} Z) Z Z^\tran] $ is diagonal. Since $\Delta$ is also diagonal by
definition, it follows that $\Lambda_o$ must be diagonal and consequently the spectral value decomposition of $\Sigma_o$ is given
by $\Sigma_o = P \Lambda_o P^\tran$.  Also, $\Sigma =\Sigma_o^{1/2} \Gamma \Sigma_o^{1/2} =
P \Lambda_o^{1/2}P^\tran P \Delta P^\tran P \Lambda_o^{1/2}P^\tran = P \Lambda P^\tran$, where $\Lambda = \Lambda_o^{1/2} \Delta \Lambda_o^{1/2}$.
Hence, the columns of $P$ correspond to both the eigenvectors of $\Sigma_o $ and the eigenvectors of $\Sigma$. \\

\noindent
\textbf{Part b:} The formula for $\lambda_j, j=1, \ldots, q$ given in \eqref{eq:Lambda} follows by pre- and post-multiply \eqref{eq:T3} by $\Lambda_o^{1/2}$  and then
extracting the diagonal elements. It is convenient for the following proofs to express \eqref{eq:Lambda} as 
 \begin{equation} \label{eq:T5} 
1 =  (1-\gamma) \alpha_i + \gamma/\lambda_i, \ i = 1, \ldots q,
\end{equation}
where $\alpha_i = E[u(\sum_{k=1}^q \beta_k Z_k^2) \beta_j Z_i^2]$  and $\beta_k = \lambda_{o,k}/\lambda_k$. \\

\noindent
\textbf{Part c:} Next, we show the eigenvalues associated with the eigenvectors of $\Sigma_o$ and $\Sigma$ have the same order. That is, for 
$\Lambda_o = diag\{\lambda_{o,1}, \ldots, \lambda_{o,q}\}$ and $\Lambda = diag\{\lambda_1, \ldots, \lambda_q\}$, $\lambda_{o,i} \ge \lambda_{o,i+1}
\Rightarrow \lambda_i \ge \lambda_{i+1}$. We also show  $\lambda_{o,i} > \lambda_{o,i+1} \Rightarrow \lambda_i > \lambda_{i+1}$. 
If $\beta_i \le \beta_{i+1}$, then this result readily follows. Thus, we only need to  consider the case for which $\beta_i > \beta_{i+1}$. For this case, we
show $\alpha_i > \alpha_{i+1}$, and so it follows that  $\gamma/\lambda_i < \gamma/\lambda_{i+1}$ or  $\lambda_i > \lambda_{i+1}$. 
To show $\alpha_i > \alpha_{i+1}$, recall $\psi(s) = s u(s)$ is increasing and $u(s)$ is decreasing in $s \ge 0$, and so $\psi\left(\sum_{k=1}^q \beta_k Z_k^2\right)$ 
is increasing and $u\left(\sum_{k=1}^q \beta_k Z_k^2\right)$ is decreasing in $\beta_i  > 0 $ when holding $\beta_j$ fixed for $j \ne i$.  
Also,  $\beta_i Z_i^2/\sum_{k=1}^q \beta_k Z_k^2$ is strictly increasing in $\beta_i  > 0 $ when holding $\beta_j$ fixed for $j \ne i$. This  implies
 \begin{eqnarray*}
 u\left(\sum_{k=1}^q \beta_k Z_k^2\right) \beta_i Z^2_i & = & \psi\left(\sum_{k=1}^q \beta_k Z_k^2\right)  
 \left\{ \frac{\beta_i Z_i^2}{\sum_{k=1}^q \beta_k Z_k^2} \right\} \\
 & > & \psi\left(\beta_{i+1} Z_i^2 + \beta_{i+1} Z_{i+1}^2 +s_i \right)  \left\{\frac{\beta_{i+1}Z^2_i}{\beta_{i+1} Z_i^2 + \beta_{i+1} Z_{i+1}^2 +s_i} \right\} \\
&  = &  u\left(\beta_{i+1} Z_i^2 + \beta_{i+1} Z_{i+1}^2 +s_i \right) \beta_{i+1}Z^2_i \\
& \ge & u\left(\beta_{i+1} Z_i^2 + \beta_{i} Z_{i+1}^2 +s_i \right) \beta_{i+1}Z^2_i ,
\end{eqnarray*}
 where $s_{i} = \sum_{k \in \kappa_i} \beta_k Z_k^2$ with $\kappa_i = \{k \ne {i,i+1}\}$. is defined as in \eqref{eq:T6}. Finally, since $Z \sim P Z$ for any permuation $P$, we obtain
 \[\alpha_i  >  E\left[u\left(\beta_{i+1} Z_i^2 + \beta_{i} Z_{i+1}^2 +s_i \right) \beta_{i+1}Z^2_i\right]  =     
 E\left[u\left(\beta_{i+1} Z_{i+1}^2 + \beta_{i} Z_{i}^2 +s_i \right) \beta_{i+1}Z^2_{i+1}\right]  = \alpha_{i+1}. \]

\noindent
\textbf{Part d:} Finally, we show the multiplicities of the eigenvalues of $\Sigma_o$ and $\Sigma$ are the same. That is, $\lambda_{o,i} = \lambda_{o,i+1} \Leftrightarrow
\lambda_i = \lambda_{i+1}$. Suppose $\lambda_{o,i} = \lambda_{o,i+1} \equiv \lambda_o$, then 
\begin{equation} \label{eq:T6} 
\Lambda_2 = diag\{\lambda_i,\lambda_{i+1}\} =  
(1-\gamma) \lambda_o \ E[u(\lambda_o Z_{(2)}^\tran \Lambda_2^{-1} Z_{(2)} +s_i) Z_{(2)} Z_{(2)}^\tran] + \gamma I_2, 
\end{equation}
where $Z_{(2)} = (Z_i, Z_{i+1})^\tran$ and $s_{i}$ is defined as above. 
Since $Z_{(2)} \sim Q_2Z_{(2)}$ for any orthogonal $Q_2$ of order $2$, it follows that $\Lambda_2$ can be replaced by  
$\Gamma_2 = Q_2 \Lambda_2 Q_2^\tran$ in \eqref{eq:T6}.
However, we know that $\Sigma$ and hence $\Gamma_2$ is uniquely defined, and so  $\Lambda_2 =  Q_2 \Lambda_2 Q_2^\tran$ for any
orthogonal $Q_2$, which implies  $\Lambda_2 \propto I_2$ or $\lambda_i = \lambda_{i+1}$. 
The other direction, i.e.\ $\lambda_i = \lambda_{i+1} \Rightarrow \lambda_{o,i} = \lambda_{o,i+1}$, follows from the result that if  $\lambda_{o,i} > \lambda_{o,i+1}$ then 
$\lambda_i > \lambda_{i+1}$, which is established above.  So, if $\lambda_i = \lambda_{i+1}$ but $\lambda_{o,i} \ne \lambda_{o,i+1}$, then either $\lambda_i > \lambda_{i+1}$ or $\lambda_i < \lambda_{i+1}$, depending on whether $\lambda_{o,i} >  \lambda_{o,i+1}$ or $\lambda_{o,i} < \lambda_{o,i+1}$ respectively, a contradiction. Hence,  $\lambda_i = \lambda_{i+1} \Rightarrow \lambda_{o,i} = \lambda_{o,i+1}$.  
 \subsection*{\small \emph{Proof of Theorem \ref{Thrm:moreS}}}
 If $\lambda_{o,i} = \lambda_{o,i+1}$, then $\lambda_{i} = \lambda_{i+1}$ and so it readily follows 
${\lambda_i}/{\lambda_{i+1}} = {\lambda_{v,i}}/{\lambda_{v, i+1}} = {\lambda_{o,i}}/{\lambda_{o, i+1}} = 1.$ 
 So, we only need to prove strict inequality holds when $\lambda_{o,i} > \lambda_{o,i+1}$, which we hereafter assume  holds. \\
 
 \noindent
 \textbf{Part a:} We first note that $(\lambda_i-\gamma)/(\lambda_{i+1}-\gamma)$ is strictly increasing in $\gamma$ on $0 \le \gamma \le 1$. Thus,
$ {\lambda_i}/{\lambda_{i+1}} < (\lambda_i-\gamma)/(\lambda_{i+1}-\gamma) = {\lambda_{v,i}}/{\lambda_{v,i+1}}$. \\

 \noindent
 \textbf{Part b:} Next, we show ${\lambda_i}/{\lambda_{i+1}} < {\lambda_{o,i}}/{\lambda_{o, i+1}}$, which is equivalent to 
$\beta_i={\lambda_{o,i}}/{\lambda_i} > {\lambda_{o, i+1}}/{\lambda_{i+1}}=\beta_{i+1}$. Since $\lambda_i>\lambda_{i+1}$, it follows that $\alpha_i>\alpha_{i+1}$, where $\alpha_i$ is defined as in \eqref{eq:T5}. It was shown within Part \textbf{c} of the proof of Theorem \ref{Thrm:eigen} that $\beta_i > \beta_{i+1}
\Rightarrow \alpha_i > \alpha_{i+1}$. Analogous arguments give $\beta_i \le \beta_{i+1} \Rightarrow \alpha_i \le \alpha_{i+1}$, a contradiction. Hence 
$\beta_i > \beta_{i+1}$. \\

\noindent
\textbf{Part c:} Finally, we show $\lambda_{v,i}/\lambda_{v, i+1}<\lambda_{o,i}/\lambda_{o,i+1}$, which is equivalent to showing $\beta_{v,i} = 
{\lambda_{o,i}}/{\lambda_{v,i}} > {\lambda_{v,i+1}}/{\lambda_{v, i+1}}=\beta_{v,i+1}$. 
From \eqref{eq:Lambda}, with $W = \sum_{k=1}^q\beta_k Z_k^2$, we have
\[
\beta_{v,i}=\lambda_{o,i}E[u(W)Z_i^2] \Longrightarrow 1=\beta_{v,i}E[u(W)Z_i^2],\quad i=1,\cdots,q.
\]
So showing $\beta_{v,i} > \beta_{v,i+1}$ is equivalent to showing $E[u(W) Z_i^2] < E[u(W)Z_{i+1}^2]$. 
 To do so, recall that the distribution of $Z$ is invariant under permutations. This gives the identity
\begin{equation} \label{eq:C}
E[u(W)Z_i^2] - E[u(W)Z_{i+1}^2]  = E[(Z_i^2-Z_{i+1}^2)\{u(W)-u(W+W_{i,i+1})\}I_{\mathcal{C}}(Z)],
\end{equation}
where $\mathcal{C}=\{z\in\mathbb{R}^q: |z_i|>|z_{i+1}|\}$ and 
\[
W_{i,i+1}=\beta_i Z_{i+1}^2+ \beta_{i+1} Z_i^2 - \beta_i Z_i^2 - \beta_{i+1} Z_{i+1}^2 = (Z_i^2 - Z_{i+1}^2)(\beta_{i+1}-\beta_i).
\]
It was established in Part \textbf{b} that $\beta_i > \beta_{i+1}$ and so $W_{i,i+1} < 0$ on $\mathcal{C}$. Since $u(\cdot)$ is a decreasing function, 
it follows that the right hand side of \eqref{eq:C} is positive and hence $E[u(W)Z_i^2] < E [u(W)Z_{i+1}^2]$. \\

\noindent
\textbf{Part d:} To complete the proof of Theorem \ref{Thrm:moreS}, we need to show $\lambda_i/\lambda_j$ 
is a decreasing function of $\gamma$ when $i < j$. Let $H(\gamma)  = \log(\lambda_i/\lambda_j)$. Using \eqref{eq:T5}, a straightforward calculations gives 
the derivative $H'(\gamma) = (\alpha_j - \alpha_i) + (\lambda_j -\lambda_i)/(\lambda_i \lambda_j)$. Since $\lambda_j < \lambda_i$ and from \eqref{eq:T5}
is follows that $\alpha_j < \alpha_i$, we have $H'(\gamma) < 0$. 

 \subsection*{\small \emph{Proof of Theorem \ref{Thrm:bdbeta}}}  
The case $0 \le \beta < 1$ has already been considered in detail in Section \ref{Sec:BP}. Consider the case $1 \le \beta < q$.
For $Y = \{ y_1, \ldots, y_m\}$ and $Z= X \cup Y$, we first show under the conditions of the theorem that $\widehat{\Sigma}_{\beta,\gamma}(Z)$ exists.
By condition A in \cite{Ollila-Tyler:2014}, existence is assured if $\# \{z_i \in \mV\}/(n+m) < dim(\mV)/\beta$ holds for any subspace $\mV$ of 
$\R^q, 1 \le dim(\mV) < q$. Since $X$ is in general position, at most $d = dim(\mV)$ sample points from $X$ can lie in $\mV$. This implies that at most
$d +m$ elements of $Z$ lies in $\mV$. Now $(d+m)/(n+m) < d/\beta$ holds for $d = 1, \ldots, q-1$ if it holds for $d=1$, i.e.\ 
$\widehat{\Sigma}_{\beta,\gamma}(Z)$ exists if $(1+m)/(n+m) < 1/\beta$ or $\epsilon_m = m/(n+m) < \epsilon^*(n-\beta)/\{(n-1)\beta\} $. 

To complete the proof, we show that for $\epsilon_m < (n-\beta)/\{(n-1)\beta\}$ the largest and smallest eigenvalues of $\widehat{\Sigma}_{\beta,\gamma}(X)$ 
are bounded above and away from zero respectively,  i.e.\ that $\widehat{\Sigma}_{\beta,\gamma}(X)$ can not breakdown under $\epsilon_m$ contamination. 
Now $\widehat{\Sigma}_{\beta,\gamma}(X \cup Y) = \widehat{\Sigma}_{\beta,\gamma}(X \cup \Theta)$ where 
$\Theta = \{ \theta_1, \ldots, \theta_m\}$ where \mbox{$\theta_i = y_i/\|y_i\|$}  for $i = 1, \ldots, m$, with $\Theta$ being a compact set.
So if breakdown occurs when $\epsilon_m < \epsilon^*$ then there exists sequence $\Theta_k \to \Theta$ for which the largest eigenvalue of 
$\widehat{\Sigma}_{\beta,\gamma}(X \cup \Theta_k)$ goes to infinity and/or its smallest eigenvalue goes to zero. However, since 
$\widehat{\Sigma}_{\beta,\gamma}(X \cup \Theta)$ is a continuous function of $\Theta$, this contradicts 
$\widehat{\Sigma}_{\beta,\gamma}(X \cup \Theta_k) \to \widehat{\Sigma}_{\beta,\gamma}(X \cup \Theta) > 0$, and hence breakdown cannot occur.
The results for $\widehat{V}_{\beta,\gamma}(X) = (\widehat{\Sigma}_{\beta,\gamma}(X) - \gamma I)/(1 - \gamma)$ follows after noting that it can only
be non-singular when $n \ge q$.

 \noindent
 \subsection*{\small \emph{Proof of Theorem \ref{thm:sbd}}}  \label{HbScat}
For $z \in \R^q, z \ne 0$ and $\Sigma > 0$, define
\beq \label{eq:d}
d( z, \Sigma) = q \log\{ z^ \tran  \Sigma^{-1}  z/ z^ \tran  z\} + \log \det\{ \Sigma\}.
\end{equation}
Since $d(c z, \beta  \Sigma)  = d( z, \Sigma)$ for any $c \in \R$ and $\beta > 0$, we can presume without loss of generality that
$ z$ is on the unit sphere $\mS_{q-1} = \{ z \in \R^q: \|  z \| =1\}$, and $\lambda _{1} = q$, where 
$\lambda_1 \ge \cdots \ge \lambda_q > 0$ are the eigenvalues of $ \Sigma$. In general, for a sequence $ \Sigma_k >  0$,  express their
spectral value decompositions as  $ \Sigma_k =  P_k  \Delta_k  P_k^ \tran$, where 
$ \Delta_k = diag\{\lambda_{1,k}, \ldots,  \lambda_{q,k}\}$
with  $\lambda_{1,k} \ge \cdots \ge \lambda_{q,k} > 0$  being the eigenvalues of $ \Sigma_k$,  and
$ P_k= [  p_{1,k} \cdots   p_{q,k} ]$ being orthogonal matrices whose columns are normalized eigenvectors of  $ \Sigma_k$, with 
$ \Sigma_k  q_{j,k} = \lambda_{j,k}   q_{j,k}$ for  $j = 1, \ldots, q$.
\newtheorem*{lemmaA1}{Lemma A.1}
\begin{lemmaA1} 
Let $ z_k,  k=1, 2, \ldots$ be a sequence in $\mS_{q-1}$,  and let  $ \Sigma_k  >   0,  k  =  1, 2, \ldots$ be  a  sequence such that 
$\lambda_{1,k} = 1$ and $\lambda_{q,k} \to 0$ as $k \to \infty$.   If the sequence $ p_{q,k}^ \tran   z_k $ bounded 
away from zero for large enough $k$, then $d( z_k, \Sigma_k) \to \infty$.
\end{lemmaA1}
\begin{proof}
Express $d( z_k,  \Sigma_k) = q \log\{\sum_{j=1}^q a_{j,k}^2/\lambda_{j,k}  \} + \sum_{j=1}^q \log \lambda_{j,k}$, where 
$a_{j,k} =  p_{j,k}^ \tran  z_k$. Since $\lambda_{1,k} = 1$,  $\log\{a_{q,k}^2\}$ is bounded below and
 $\lambda_{q,k} \to 0$ as $k \to \infty$, it follows that \\[4pt] \hspace*{1cm}
$d( z_k,  \Sigma_k) \ge q \log\{a_{q,k}^2/\lambda_{q,k}  \} + (q-1) \log \lambda_{q,k}  = q \log\{a_{q,k}^2\} - \log \lambda_{q,k} 
\to \infty.$
\end{proof}
For the sample $ Z = \{ z_i \in \R^q; i  = 1,  \ldots, n+m \}$ and $ \Sigma >  0$, define
\begin{equation} \label{eq:D}
 D( Z;  \Sigma) = median \{d( z_{i}, \Sigma); i =  1, \ldots,  n+m,  z_i \ne  0\}.
 \end{equation}
 \newtheorem*{lemmaA2}{Lemma A.2}
 \begin{lemmaA2} 
Suppose the fixed sample $ X = \{ x_i \in \mS_{q-1}; i  = 1,  \ldots, n \}$ is in general position, and 
$ Y = \{ y_{i} \in \mS_{q-1}; i = 1, \ldots, m \}$ is an arbitrary sample.
For a sequence $ \Sigma_k >  0$ with $\lambda_{q,k} \to 0$, if
$m < n-2q+2$ then $\inf_{ Y}  D( X \cup  Y;  \Sigma_k) \to \infty$.
\end{lemmaA2}
\begin{proof}
For any $ x_i \in  X$, it follows from Lemma A.1 that $d( x_i,  \Sigma_k) \to \infty$ unless there is a sequence in  $k$
such that $ p_{q,k}^ \tran  x_i \to 0$ as $k \to \infty$. Since $ p_{q,k}$ lies in a compact set, it can be assumed that
$ p_{q,k} \to  p_q \in \mS_{q-1}$ and so if $ p_{q,k}^ \tran  x_i\to 0$ holds, then $ p_{q}^ \tran  x_i\ = 0$. Since $ X$ 
is in general position, $ p_{q}^ \tran  x_i\ = 0$ holds for at most $q-1$ element of $ X$.  This implies $d( x_{i}, \Sigma) \to \infty$ 
for at least $n-q+1$ sample points in $ X$. The result then follows since $  X \cup  Y$ consist of $n+m < 2(n-q+1)$ sample points.
\end{proof}
For the sample $ Z$, we can express $\widehat{ \Sigma}_R( Z)  = \arg\inf \{ D( Z;  \Sigma) ~|~   \Sigma >  0 \}$,
provided the infimum exists.  If the infimum exists and is achieved at  $\widehat{ \Sigma}_R( Z) >  0$, then it is also achieved at 
$c \widehat{\Sigma}_R( Z)$ for any $c > 0$.  All such solutions give the same value for 
$\widehat{ \Sigma}_{sh,R}( Z) = q \widehat{ \Sigma}_{sh,R}( Z)/\tr\{\widehat{ \Sigma}_{sh,R}( Z)\} $.
It may be possible though for $\widehat{ \Sigma}_R( Z)$ to not be well defined since the global minimum above may not be unique up to
scalar multiplication. In such an unlikely case, $\widehat{ \Sigma}_{sh,R}( Z)$ can be taken to be the average of all such values. Conditions
for the existence of $\widehat{ \Sigma}_R( Z)$ are given in the following lemma.
\newtheorem*{lemmaA3}{Lemma A.3} 
\begin{lemmaA3} 
Let $ X$  and $ Y$ be defined as in Lemma A.2  with  $m < n-2q+2$. For a given $ Y$, $D( X \cup  Y;  \Sigma)$ is 
bounded below over $ \Sigma >  0$. Furthermore, $\widehat{ \Sigma}_{sh}( X \cup  Y) >  0$ exists and
$\inf_{ Y} \det \{\widehat{ \Sigma}_{sh}( X \cup  Y)\} > 0$.
\end{lemmaA3}
\begin{proof}
In general $D( Z;  \Sigma) \ge \log\{\det  \Sigma\}$. So, if $D( X \cup  Y;  \Sigma)$ is not bounded below
then there exists a sequence  $ \Sigma_k >   0$ such that $D( X \cup  Y;  \Sigma_k) \to -\infty$. This implies $\det  \Sigma_k \to  0$ or
equivalently $\lambda_{q,k} \to 0$.  However, Lemma A.2 then states $D( X \cup  Y;  \Sigma_k) \to \infty$, a contradiction.
Now, since $D( X \cup   Y;  \Sigma)$ is also continuous over $ \Sigma >   0$, it follows 
$\widehat{ \Sigma}_{sh,R}( X \cup  Y) >  0$ exists. Finally, if $\inf_{ Y} \det \{\widehat{ \Sigma}_{sh,R}( X \cup  Y)\} = 0$, then
Lemma A.2 is again contradicted. 
\end{proof}
The breakdown point for $\widehat{ \Sigma}_{sh,R}( X)$ given in Theorem \ref{thm:sbd} is an immediate consequence of 
Lemmas A.2 and A.3, since for $m < n-2q+2$,  $\widehat{ \Sigma}_{sh,R}( X \cup  Y)$ exist with its largest eigenvalue equal to $q$
by definition and its  smallest eigenvalue bounded away from zero for all $Y$. The breakdown points of $\widehat{ \Sigma}_{sh,\widetilde{\beta}}(X)$ and 
$\widehat{ \Sigma}_{sh,\widetilde{\beta}}(X)$ also follows since under $X \cup Y$ and for $m < n-2q+2$ we know they exist and their largest eigenvalues 
are again equal to $q$. Finally, we know that under $X \cup Y$ and for $m < n-2q +2$ their smallest eigenvalue is bounded away from zero over all $Y$.
Otherwise $\inf_Y D( X \cup   Y;  \widehat{\Sigma}_\beta)$ would not be bounded above for all $0 \le \beta <  q$ which contradicts the breakdown point
being one for any $0 < \beta < 1$.

\end{document}